\documentclass[longauth]{aa} 

\usepackage{graphicx}
\usepackage{txfonts}
\usepackage{xcolor}
\begin{document} 

   \title{CON-quest}

   \subtitle{Searching for the most obscured galaxy nuclei}

   \author{N.~Falstad\inst{1}
     \and
     S.~Aalto\inst{1}
     \and
     S.~K{\"o}nig\inst{1}
     \and
     K.~Onishi\inst{1}
     \and
     S.~Muller\inst{1}
     \and
     M.~Gorski \inst{1}
     \and
     M.~Sato \inst{1}
     \and
     F.~Stanley \inst{1}
     \and
     F.~Combes \inst{2}
     \and
     E.~Gonz{\'a}lez-Alfonso\inst{3}
     \and
     J.~G.~Mangum\inst{4}
     \and
     A.~S.~Evans\inst{4,5}
     \and
     L.~Barcos-Mu{\~n}oz\inst{4}
     \and
     G.~C.~Privon\inst{4}
     \and
     S.~T.~Linden\inst{5}
     \and
     T.~D{\'i}az-Santos\inst{6,24,25}
     \and
     S.~Mart{\'i}n\inst{7,8}
     \and
     K.~Sakamoto\inst{9}
     \and
     N.~Harada\inst{9,18}
     \and
     G.~A.~Fuller\inst{10,26}
     \and
     J.~S.~Gallagher\inst{11}
     \and
     P.~P.~van~der~Werf\inst{12}
     \and
     S.~Viti\inst{12,13}
     \and
     T.~R.~Greve\inst{14,13}
     \and
     S.~Garc{\'i}a-Burillo\inst{15}
     \and
     C.~Henkel\inst{16,17}
     \and
     M.~Imanishi\inst{18}
     \and
     T.~Izumi\inst{18}
     \and
     Y.~Nishimura\inst{19,18}
     \and
     C.~Ricci\inst{6,21,22}
     \and
     S.~M{\"u}hle\inst{23}
   }

   \institute{Department of Space, Earth and Environment, Chalmers University of Technology, Onsala Space Observatory,
     439 92 Onsala, Sweden \\
     \email{niklas.falstad@chalmers.se}
     \and
     Observatoire de Paris, LERMA, College de France, CNRS, PSL Univ., Sorbonne University, UPMC, Paris, France
     \and
     Universidad de Alcal{\'a}, Departamento de F{\'i}sica y Matem{\'a}ticas, Campus Universitario, E-28871 Alcal{\'a} de Henares, Madrid, Spain
     \and
     National Radio Astronomy Observatory, 520 Edgemont Road, Charlottesville, VA 22903, USA
     \and
     Department of Astronomy, 530 McCormick Road, University of Virginia, Charlottesville, VA 22904, USA
     \and
     N{\'u}cleo de Astronom{\'i}a de la Facultad de Ingenier{\'i}a, Universidad Diego Portales, Av. Ej{\'e}rcito Libertador 441, Santiago, Chile
     \and
     European Southern Observatory, Alonso de C{\'o}rdova 3107, Vitacura  763 0355, Santiago, Chile
     \and
     Joint ALMA Observatory, Alonso de C{\'o}rdova 3107, Vitacura 763 0355, Santiago, Chile
     \and
     Institute of Astronomy and Astrophysics, Academia Sinica, 11F of AS/NTU Astronomy-Mathematics Building, No.1, Sec. 4, Roosevelt Rd, Taipei 10617, Taiwan, R.O.C.
     \and
     Jodrell Bank Centre for Astrophysics, Department of Physics \& Astronomy, School of Natural Sciences, The University of Manchester, M13\,9PL, UK
     \and
     Department of Astronomy, University of Wisconsin-Madison, 5534 Sterling, 475 North Charter Street, Madison WI 53706, USA
     \and
     Leiden Observatory, Leiden University, P.O.\ Box 9513, NL-2300 RA Leiden, The Netherlands
     \and
     Department of Physics and Astronomy, University College London, Gower Street, London WC1E\,6BT, UK
     \and
     Cosmic Dawn Center (DAWN), DTU-Space, Technical University of Denmark, Elektrovej 327, DK-2800 Kgs. Lyngby, Denmark
     \and
     Observatorio de Madrid, OAN-IGN, Alfonso XII, 3, E-28014-Madrid, Spain
     \and
     Max-Planck-Institut f{\"u}r Radioastronomie, Auf dem H{\"u}gel 69, 53121, Bonn, Germany
     \and
     Astron. Dept., King Abdulaziz University, P.O. Box 80203, 21589 Jeddah, Saudi Arabia
     \and
     National Astronomical Observatory of Japan, National Institutes of Natural Sciences (NINS), 2-21-1 Osawa, Mitaka, Tokyo 181–8588, Japan
     \and
     Institute of Astronomy, The University of Tokyo, 2-21-1, Osawa, Mitaka, Tokyo 181-0015, Japan
     \and
     Chile Observatory, National Astronomical Observatory of Japan, 2-21-1, Osawa, Mitaka, Tokyo 181-8588, Japan
     \and
     Kavli Institute for Astronomy and Astrophysics, Peking University, Beijing 100871, China
     \and
     George Mason University, Department of Physics \& Astronomy, MS 3F3, 4400 University Drive, Fairfax, VA 22030, USA
     \and
     Argelander Institut f{\"u}r Astronomie, Universit{\"a}t Bonn, Auf dem H{\"u}gel 71, 53121 Bonn, Germany
     \and
     Chinese Academy of Sciences South America Center for Astronomy (CASSACA), National Astronomical Observatories, CAS, Beijing 100101, China
     \and
     Institute of Astrophysics, Foundation for Research and Technology--Hellas (FORTH), Heraklion, GR-70013, Greece
     \and Instituto de Astrof\'isica de Andalucia (CSIC), Glorieta de al Astronomia s/n E-18008, Granada, Spain
 }
   \date{}

  \abstract
      {Some luminous and ultraluminous infrared galaxies (LIRGs and ULIRGs) host extremely compact ($r<100$~pc) and dusty nuclei. The high extinction associated with large column densities of gas and dust toward these objects render them hard to detect at many wavelengths. The intense infrared radiation arising from warm dust in these sources can provide a significant fraction of the bolometric luminosity of the galaxy and is prone to excite vibrational levels of molecules such as HCN. This results in emission from the rotational transitions of vibrationally excited HCN (HCN-vib), with the brightest emission found in compact obscured nuclei (CONs; $\Sigma_{\mathrm{HCN-vib}}> 1$~L$_{\sun}$\,pc$^{-2}$ in the  $J=3\text{--}2$ transition). However, there have been no systematic searches for CONs, and it is unknown how common they are.}
   {We aim to establish how common CONs are in the local Universe ($z<0.08$), and whether their prevalence depends on the luminosity or other properties of the host galaxy.}
   {We have conducted an Atacama Large Millimeter/submillimeter Array (ALMA) survey of the rotational $J=3\text{--}2$ transition of HCN-vib in a volume-limited sample of $46$ far-infrared luminous galaxies.}
   {Compact obscured nuclei are identified in $38^{+18}_{-13}\%$ of the ULIRGs, $21^{+12}_{-6}\%$ of the LIRGs, and $0^{+9}_{-0}\%$ of the lower luminosity galaxies. We find no dependence on the inclination of the host galaxy, but strong evidence of lower IRAS $25$ to $60$~$\mu$m flux density ratios ($f_{25}/f_{60}$) in CONs (with the exception of one galaxy, NGC~4418) compared to the rest of the sample. Furthermore, we find that CONs have stronger silicate features ($s_{9.7\mathrm{\mu m}}$) but similar polycyclic aromatic hydrocarbon equivalent widths (EQW$_{6.2\mathrm{\mu m}}$) compared to other galaxies. Besides signatures of molecular inflows seen in the far infrared in most CONs, submillimeter observations also reveal compact, often collimated, outflows.}
   {In the local Universe, CONs are primarily found in (U)LIRGs, in which they are remarkably common. As such systems are often highly disturbed, inclinations are difficult to estimate, and high resolution continuum observations of the individual nuclei are required to determine if the CON phenomenon is related to the inclinations of the nuclear disks. Further studies of the in- and outflow properties of CONs should also be conducted in order to investigate how these are connected to each other and to the CON phenomenon. The lower $f_{25}/f_{60}$ ratios in CONs as well as the results for the mid-infrared diagnostics investigated (EQW$_{6.2\mathrm{\mu m}}$ and $s_{9.7\mathrm{\mu m}}$) are consistent with the notion that large dust columns gradually shift the radiation from the hot nucleus to longer wavelengths, making the mid- and far-infrared “photospheres” significantly cooler than the interior regions. Finally, to assess the importance of CONs in the context of galaxy evolution, it is necessary to extend this study to higher redshifts where (U)LIRGs are more common.}


   \maketitle
%

   \section{Introduction}\label{sec:introduction}
   Over the last decade, it has been found that some luminous ($10^{11}$~L$_{\sun} < L_{\mathrm{IR}}(8-1000\mathrm{\mu m}) < 10^{12}$~L$_{\sun}$) and ultraluminous ($L_{\mathrm{IR}} > 10^{12}$~L$_{\sun}$) infrared galaxies (LIRGs and ULIRGs) in the local Universe exhibit emission from rotational lines of HCN in its vibrationally excited $v_{2}=1$ state \citep[hereafter HCN-vib; e.g.,][]{sak10,ima13,aal15b,mar16}. The brightest emission has been found in compact obscured nuclei (CONs; $\Sigma_{\mathrm{HCN-vib}}> 1$~L$_{\sun}$\,pc$^{-2}$ in the $J=3\text{--}2$ transition, $r>5$~pc) whose dusty cores may be optically thick up to millimeter wavelengths \citep[e.g.,][]{sak13,mar16,sak17,sco17,aal19}. In these regions, trapping of continuum photons by dust creates a ``greenhouse'' effect that may increase the central dust temperatures to several hundreds of Kelvin \citep{gon19}. These high temperatures give rise to an intense mid-infrared radiation field that is able to populate the vibrational states of molecules such as HCN. With sizes on the order of $10$ to $100$~pc and dust temperatures around $100$~K at the far-infrared photosphere, CONs may be able to supply a significant fraction of the total infrared luminosity of their host galaxies \citep[e.g.,][]{gon12,fal15}. Due to the obscured nature of their nuclei, it is still unclear what the ultimate embedded source of the high luminosity in CONs is; it may be an accreting supermassive black hole (SMBH) in an active galactic nucleus (AGN), a nuclear starburst, or a combination of both. If CONs are mainly powered by hidden AGN activity, they may represent a phase of rapid accretion onto the SMBH, almost completely surrounded by high column densities of obscuring material, following a merger or interaction event \citep{koc15,ric17,ble18,boe20}. Regardless of the exact nature of the hidden power source, studies of these objects could help us understand growth processes in galaxy nuclei, as well as the relations between black hole mass and bulge properties \citep[e.g.,][]{mag98,kor13}. An interesting comparison may be made with the compact star-forming galaxies found at redshifts $2-3$ \citep[e.g.,][]{bar13}. These are heavily dust obscured \citep{bar14} galaxies with a high incidence of AGN activity \citep{bar14,koc17} and are thought to form through the dissipative contraction of gas-rich galaxies, triggering a central starburst rapidly growing the stellar bulge \citep[e.g.,][]{zol15,koc17,tac18}.

   An important way in which the intense nuclear activity may influence the host galaxy is by mechanical feedback in the form of winds and outflows. A large fraction of ULIRGs are observed to have wide angle molecular outflows detectable using the median velocities of far-infrared OH absorptions \citep{vei13}. These outflow signatures have never been observed in CONs, which instead often show signatures of inflowing gas in the far-infrared OH lines \citep{fal19}. However, at least some of the CONs host compact and collimated outflows observable at longer wavelengths using spectral lines of, for example, CO and HCN \citep[e.g.,][]{bar18,fal18,fal19}. It is still unclear whether such compact outflows can be seen in all CONs and whether they are connected to the kiloparsec-scale dust features seen in optical images \citep[e.g.,][]{fal18,aal19}, either as the base of continuous outflows or as the most recent outbursts in recurring cycles of outflow events. 

   A major problem when studying the central regions of CONs is that the high column densities preclude observations at wavelengths up to at least the far-infrared due to dust absorption. Some objects are opaque at almost all wavelengths with only a narrow range between the submillimeter and radio that is optically thin \citep[e.g., Arp~220;][]{bar15}. Although observed at (sub)millimeter wavelengths where the dust emission may become optically thin, interpretation of molecular spectral line emission can be difficult. The radiative transfer of traditional tracers of dense molecular gas such as HCN and HCO$^{+}$ is complicated by absorption due to large columns of cooler foreground gas along the line of sight to the nucleus \citep[e.g.,][]{aal15b,mar16,ima16a}.

   In these situations, HCN-vib provides a tool to probe deeper into the nucleus \citep[e.g.,][]{mar16,aal19}, but it can also be used as a survey tool to search for obscured activity \citep[e.g.,][]{aal19}. The rotational transitions of HCN-vib occur within the vibrationally excited $v_{2}=1$ state which has an energy above ground of $E/k=1024$~K. This state is doubly degenerate, so that the rotational states are split into two levels, f and e. The $J=3\text{--}2$ $v_{2}=1f$ transition which we use in this paper occurs at a frequency $267.199$~GHz. The critical density of the $v_{2}=1\text{--}0$ transition is $\sim5\times 10^{11}$~cm$^{-3}$, high enough for collisional excitation to be unlikely. Instead, the vibrational state is likely populated through radiative excitation directly from the ground state by $14$~$\mu$m photons \citep{ziu86}. For the state to be efficiently populated in this way, radiation with a brightness temperature in excess of $100$~K at $14$~$\mu$m is required \citep{aal15b}. Since the first extragalactic detection by \citet{sak10}, different HCN-vib transitions have been observed in many other galaxies \citep[e.g.,][]{ima13,ima16a,ima16b,ima18,aal15a,aal15b,mar16}. Efforts have been made to make inventories of these existing extragalactic detections. For example, in a sample of nine galaxies, mainly (U)LIRGs, by \citet{aal15b}, HCN-vib lines were detected in eight. Five of these could be classified as CONs with the definition used at the time. A follow-up study by \citet{fal19} found $19$ (U)LIRGs with existing observations covering the HCN-vib frequencies. Of these, $13$ were detected and $7$, including the $5$ found by \citet{aal15b}, could be classified as CONs. However, there have been no systematic searches for CONs, and an understanding of the prevalence of CONs will help assess their importance to the galaxy evolution process.

 In this paper, we present the first results of a volume limited survey of infrared luminous galaxies, named CON-quest, with the aim to establish how common CONs are in the local ($z<0.08$) Universe. We introduce our selection criterion in Sect. \ref{sec:definition} and the sample in Sect. \ref{sec:sample}, describe the new observations and data reduction in Sect. \ref{sec:observations}, and present the results in Sect. \ref{sec:results}. In Sect. \ref{sec:discussion} we discuss the results before summarizing our conclusions in Sect. \ref{sec:conclusions}. 

\section{How to find compact obscured nuclei}\label{sec:definition}

Compact obscured nuclei are characterized by high column densities of warm dust and gas within relatively small distances of their centers, typically much smaller than the size of central molecular zones which have radii of several $100$~pc \citep[e.g.,][]{isr20}. In principle, this means that properties such as the column density and temperature of obscuring material, and the physical size of the nucleus must be determined to identify CONs. Each source would then require detailed radiative transfer modeling and excitation analysis \citep[e.g.,][]{gon12}, or high-resolution multifrequency continuum observations \citep[e.g.,][]{aal19}. However, with the discovery of bright HCN-vib emission from known CONs, a new tool to be used in surveys such as this one was provided. 

  Previous studies of CONs have often focused on the ratio between the HCN-vib luminosity and the total infrared luminosity of the galaxy, with CONs generally having $L_{\mathrm{HCN-vib}}/L_{\mathrm{IR}}>10^{-8}$ \citep[e.g.,][]{aal15b,fal19}. While this criterion is easy to use, it has a significant drawback in that it includes both emission originating mainly in the nucleus ($L_{\mathrm{HCN-vib}}$) and emission that is likely to have a substantial contribution from the entire host galaxy ($L_{\mathrm{IR}}$). This means that it may miss CONs in systems where the infrared emitting region is spatially extended or where multiple nuclei are present. Instead of using this criterion, we identify as CONs galaxies with $\Sigma_{\mathrm{HCN-vib}}> 1$~L$_{\sun}$\,pc$^{-2}$ in the $J=3\text{--}2$ transition over a region with a radius of at least $5$ pc. To avoid issues related to line blending, we use the size of the $1.1$~mm continuum when computing the HCN-vib surface brightness. In general, this is slightly larger than the size of the HCN-vib emitting region but a lower limit to the size of the far-infrared photosphere of the source \citep{gon19}. The impact of this on $\Sigma_{\mathrm{HCN-vib}}$ can be estimated by comparing the extents of the continuum and HCN-vib emission in sources where the latter is resolved and not affected by line blending. On doing so, we find that the surface brightness of HCN-vib is generally underestimated by less than a factor of two ($1.8$ in ESO 320-G030, $1.5$ in IRAS 17578-0400, and $1.3$ in ESO 173-G015). 

Our new criterion, just as the previously used one, is still purely empirical and based on the properties of known CONs and less obscured sources. For example, the CONs Arp 220~W and Zw~049.057, which both have column densities $N_{\mathrm{H_{2}}}\gtrsim5 \times 10^{24}$~cm$^{-2}$ toward the central ${\sim}50$~pc \citep{gon12,fal15}, have $\Sigma_{\mathrm{HCN-vib}}>3$~L$_{\sun}$\,pc$^{-2}$ \citep[][; this work]{aal15b,mar16} while Mrk 231, which is generally not considered a CON due to its lower column densities ($N_{\mathrm{H_{2}}}{\sim}10^{24}$~cm$^{-2}$) on similar scales \citep{gon14}, has $\Sigma_{\mathrm{HCN-vib}}{\sim}0.3$~L$_{\sun}$\,pc$^{-2}$ \citep{aal15a}. For further comparison, our limit is approximately two orders of magnitude larger than the HCN-vib surface brightness seen in the super star clusters of NGC 253 \citep{kri20}.

Due to the complex dependence of the HCN-vib emission on other properties of the source \citep{gon19}, a measured HCN-vib surface brightness can not be directly translated into physical properties such as the column density and size of an observed source. However, the examples in the previous paragraph may give some indication of the compactness and amount of obscuration in sources selected by our criterion. Furthermore, some properties can be estimated from the recent work by \citet{gon19} who model the continuum and HCN-vib emission from buried nuclear regions of galaxies. The models employ a spherically symmetric approach, simulating either an AGN or a nuclear starburst as the heating source and exploring the ranges $N_{\mathrm{H_{2}}}=10^{23}\text{--}10^{25}$~cm$^{-2}$ for the H$_{2}$ column density and $\Sigma_{\mathrm{IR}}=1.4\times 10^{7}\text{--}2.2\times 10^{8}$~L$_{\sun}$\,pc$^{-2}$ for the surface brightness. For the typical luminosity of a LIRG or ULIRG in our sample, this corresponds to radii of $\sim17\text{--}70$~pc or $\sim50\text{--}200$~pc, respectively, assuming that all of the infrared luminosity is produced in a single component. In these simplified models, $\Sigma_{\mathrm{HCN-vib}}> 1$~L$_{\sun}$\,pc$^{-2}$ in the $J=3\text{--}2$ transition can be achieved in the whole range of explored surface brightnesses and with both types of heating sources, as long as the column density exceeds $N_{\mathrm{H_{2}}}>10^{24}$~cm$^{-2}$, corresponding to an optical depth of $\sim1.5$ at $100$~$\mu$m. We note that this criterion, and in fact any criterion based on the HCN-vib luminosity, assumes that, apart from large columns of obscuring material, a CON also has a power source that is strong enough to produce the mid-infrared radiation field required to excite HCN-vib.

 \section{Sample}\label{sec:sample}
 As mentioned in Sect. \ref{sec:introduction}, no systematic searches for CONs have been carried out and only seven have been found so far: two in ULIRGs, five in LIRGs and none in sub-LIRGs \citep[$10^{10}$~L$_{\sun} \leq L_{\mathrm{IR}}<10^{11}$~L$_{\sun}$;][]{aal15b,fal19}. Due to the limited statistical data, it is still uncertain how common the CONs are, and how their prevalence depends on luminosity. As a first step to remedy this, we have compiled a sample of far-infrared luminous galaxies drawn from the IRAS revised bright galaxy sample \citep[RBGS;][]{san03}. The RBGS is a complete sample of extragalactic objects with IRAS $60$~$\mu$m flux densities greater than $5.24$~Jy, covering the entire sky surveyed by IRAS at Galactic latitudes $\lvert b\rvert>5$\degr. Our CON-quest sample was selected from the RBGS based on the far-infrared luminosities, declinations, and distances listed in Table 1 of \citet{san03} using the following far infrared luminosity, $L_{\mathrm{FIR}}(40-400\mathrm{\mu m})$, criteria: 
\begin{itemize}
\item $10^{12}$~L$_{\sun} \leq L_{\mathrm{FIR}}$, $\delta <30\degr$, $D<330$~Mpc
 
\item $10^{11}$~L$_{\sun} \leq L_{\mathrm{FIR}} < 10^{12}$~L$_{\sun}$, $\delta <15\degr$, $D<76$~Mpc
 
\item $10^{10}$~L$_{\sun} \leq L_{\mathrm{FIR}} < 10^{11}$~L$_{\sun}$, $\delta <15\degr$, $10<D<15.5$~Mpc
\end{itemize}
Declination and distance limits were set to ensure reasonable integration times ($t_{\mathrm{int}}\lesssim 2$~h) and sample sizes ($\leq20$ in each luminosity bin) to make the survey feasible. Calculating the far-infrared luminosity using the prescription in Table 1 of \citet{san96} and the relatively conservative assumptions $f_{100}=2\times f_{60}$ and $C=1.8$, we note that the full RBGS is volume limited down to the relevant far-infrared luminosities for the low-, and mid-luminosity bins. For the distance limit in the high-luminosity bin, however, the RBGS is only volume limited for $L_{\mathrm{FIR}}\gtrsim1.8\times10^{12}$~L$_{\sun}$. On comparing with the more sensitive IRAS $1.2$-Jy redshift survey \citep{str92,fis95}, we see that only one source, IRAS~06035-7102, would be added if we used the same selection criteria on that sample.

With the exception of the LIRG NGC~1068 which is in the low-luminosity bin when selecting based on far-infrared luminosity (between $40$ and $400$~$\mu$m), the high-, mid-, and low-luminosity bins contain ULIRGs, LIRGs, and sub-LIRGs, which are defined using total infrared luminosity (between $8$ and $1000$~$\mu$m), respectively, and we will refer to them by these names. In total, our criteria select $48$ systems (some of which contain multiple galaxies), eight in the ULIRG-sample and $20$ each in the LIRG and sub-LIRG samples. However, observations have not been obtained of the galaxy pairs IC 4687/6 in the LIRG sample and NGC~4568/7 in the sub-LIRG sample. A list of all observed galaxies and some of their properties is presented in Table \ref{tab:sample}.

\begin{table*}[htbp]
\caption{Sample galaxies}
\begin{center}
\begin{tabular}{lcccccccc}
\hline \hline
Name & RA & Dec & $z$ & $D_{L}$\tablefootmark{a} & $L_{\mathrm{FIR}}$\tablefootmark{b} & $L_{\mathrm{IR}}$\tablefootmark{c} & $f_{25}/f_{60}$\tablefootmark{d} & Incl.\tablefootmark{e} \\ 
 & (J2000) & (J2000) &  & (Mpc) &  ($10^{11}$ L$_{\sun}$) &($10^{11}$ L$_{\sun}$) &   & ($\degr$) \\ 
\hline
\noalign{\smallskip}
\multicolumn{8}{c}{ULIRGs} \\
\noalign{\smallskip}
\hline
\object{IRAS 17208-0014} & 17:23:21.95 & -00:17:00.9 & 0.0428 & $183 \pm 12$ & $22$ & $25 \pm 3$ & 0.05 & 90 \\ 
\object{IRAS F14348-1447} & 14:37:38.32 & -15:00:24.0 & 0.0830 & $362 \pm 24$ & $18$ & $20 \pm 3$ & 0.08 & 69 \\ 
\object{IRAS F12112+0305} & 12:13:46.08 & +02:48:41.5 & 0.0733 & $318 \pm 21$ & $17$ & $19 \pm 3$ & 0.08 & 22 \\ 
\object{IRAS 13120-5453} & 13:15:06.33 & -55:09:22.8 & 0.0311 & $135 \pm 9$ & $15$ & $18 \pm 2$ & 0.07  & 41 \\ 
\object{IRAS 09022-3615} & 09:04:12.72 & -36:27:01.3 & 0.0596 & $253 \pm 17$ & $15$ & $18 \pm 2$ & 0.10 & 37 \\ 
\object{Arp 220} & 15:34:57.27 & +23:30:10.5 & 0.0181 & $81.8 \pm 5.5$ & $14$ & $16 \pm 2$ & 0.08 & 57 \\ 
\object{IRAS F14378-3651} & 14:40:59.01 & -37:04:31.9 & 0.0682 & $295 \pm 20$ & $12$ & $14 \pm 2$ & 0.10  & $\ldots$ \\ 
\object{IRAS F22491-1808} & 22:51:49.31 & -17:52:24.0 & 0.0778 & $328 \pm 22$ & $12$ & $13 \pm 2$ & 0.10 & $\ldots$ \\
\hline
\noalign{\smallskip}
\multicolumn{8}{c}{LIRGs} \\
\noalign{\smallskip}
\hline
\object{NGC 1614} & 04:34:00.03 & -08:34:44.6 & 0.0159 & $64.1 \pm 4.3$ & $2.7$ & $4.0 \pm 0.5$ & 0.23  & 42 \\ 
\object{NGC 7469} & 23:03:15.67 & +08:52:25.3 & 0.0164 & $66.5 \pm 4.4$ & $2.5$ & $3.9 \pm 0.5$ & 0.22 & 30 \\ 
\object{NGC 3256} & 10:27:51.23 & -43:54:16.6 & 0.0094 & $36.1 \pm 2.4$ & $2.7$ & $3.7 \pm 0.5$ & 0.15 & 48 \\ 
\object{IRAS F17138-1017} & 17:16:35.70 & -10:20:38.0 & 0.0173 & $77.3 \pm 5.2$ & $1.8$ & $2.6 \pm 0.3$ & 0.14 & 71 \\ 
\object{IRAS 17578-0400} & 18:00:31.85 & -04:00:53.5 & 0.0134 & $60.0 \pm 4.1$ & $2.0$ & $2.3 \pm 0.3$ & 0.04 & 81 \\ 
\object{NGC 713}0 & 21:48:19.52 & -34:57:04.8 & 0.0163 & $68.1 \pm 4.6$ & $1.7$ & $2.3 \pm 0.3$ & 0.13 & 34 \\ 
\object{ESO 173-G015} & 13:27:23.77 & -57:29:22.1 & 0.0100 & $32.7 \pm 2.3$ & $1.8$ &  $2.2 \pm 0.3$ & 0.09 & 90 \\ 
\object{NGC 3110} & 10:04:02.12 & -06:28:29.1 & 0.0169 & $74.3 \pm 5.0$ &  $1.6$ & $2.0 \pm 0.3$ & 0.10 & 65 \\ 
\object{IC 4734} & 18:38:25.69 & -57:29:25.8 & 0.0156 & $67.8 \pm 4.5$ & $1.7$ & $1.9 \pm 0.3$ & 0.09 & 57 \\ 
\object{Zw 049.057} & 15:13:13.10 & +07:13:32.0 & 0.0130 & $60.2 \pm 4.1$ &  $1.7$ & $1.9 \pm 0.3$ & 0.04 & 44 \\ 
\object{NGC 5135} & 13:25:44.06 & -29:50:01.2 & 0.0137 & $52.6 \pm 3.5$ & $1.1$ &  $1.5 \pm 0.2$ & 0.14 & 25 \\ 
\object{ESO 221-IG10} & 13:50:56.92 & -49:03:19.7 & 0.0105 & $59.0 \pm 11.0$ & $1.0$ &  $1.5 \pm 0.6$ & 0.14 & 24 \\ 
\object{IC 5179} & 22:16:09.09 & -36:50:37.5 & 0.0113 & $47.3 \pm 3.2$ &  $1.1$ & $1.5 \pm 0.2$ & 0.12 & 62 \\ 
\object{UGC 2982} & 04:12:22.67 & +05:32:49.1 & 0.0177 & $70.6 \pm 4.7$ &  $1.0$ & $1.4 \pm 0.2$ & 0.10  & 64 \\ 
\object{ESO 286-G035} & 21:04:11.16 & -43:35:32.5 & 0.0174 & $73.2 \pm 4.9$ & $1.0$ &  $1.3 \pm 0.2$ & 0.09 & 79 \\ 
\object{NGC 4418} & 12:26:54.62 & -00:52:39.4 & 0.0073 & $33.2 \pm 2.2$ &  $1.0$ & $1.3 \pm 0.2$ & 0.22   & 68 \\ 
\object{NGC 2369} & 07:16:37.75 & -62:20:37.5 & 0.0108 & $44.6 \pm 3.0$ &  $1.0$ & $1.3 \pm 0.2$ & 0.11  & 90 \\ 
\object{NGC 5734} & 14:45:09.05 & -20:52:13.8 & 0.0136 & $60.2 \pm 4.0$ &  $1.0$ & $1.1 \pm 0.2$ & 0.09  & 58 \\ 
\object{ESO 320-G030} & 11:53:11.72 & -39:07:49.1 & 0.0103 & $36.0 \pm 2.5$ &  $1.0$ & $1.1 \pm 0.2$ & 0.07 & 64 \\ 
\hline
\noalign{\smallskip}
\multicolumn{8}{c}{sub-LIRGs} \\
\noalign{\smallskip}
\hline
\object{NGC 1068}\tablefootmark{f} & 02:42:40.77 & -00:00:47.8 & 0.0038 & $15.8 \pm 1.1$ &  $0.78$ & $2.5 \pm 0.3$ & 0.45 & 35 \\ 
\object{NGC 1808} & 05:07:42.34 & -37:30:47.0 & 0.0034 & $13.1 \pm 0.9$ &  $0.35$ & $0.56 \pm 0.08$ & 0.16  & 83 \\ 
\object{NGC 4254} & 12:18:49.63 & +14:24:59.4 & 0.0080 & $15.3 \pm 2.8$ & $0.26$ &  $0.34 \pm 0.10$ & 0.12  & 20 \\ 
\object{NGC 4303} & 12:21:54.95 & +04:28:24.9 & 0.0052 & $15.3 \pm 2.8$ & $0.23$ & $0.32 \pm 0.10$ & 0.13  & 18 \\ 
\object{NGC 660}  & 01:43:02.35 & +13:38:44.5 & 0.0028 & $12.3 \pm 0.8$ & $0.24$ & $0.31 \pm 0.04$ & 0.11  & 79 \\ 
\object{NGC 4527} & 12:34:08.50 & +02:39:13.7 & 0.0058 & $15.3 \pm 2.8$ & $0.19$ & $0.26 \pm 0.10$ & 0.11  & 81 \\ 
\object{NGC 3627} & 11:20:15.03 & +12:59:28.6 & 0.0024 & $10.0 \pm 1.1$ & $0.17$ & $0.24 \pm 0.05$ & 0.13  & 68 \\ 
\object{NGC 613}  & 01:34:18.24 & -29:25:06.6 & 0.0049 & $15.0 \pm 3.0$ & $0.17$ & $0.23 \pm 0.09$ & 0.16  & 36 \\ 
\object{NGC 4666} & 12:45:08.68 & -00:27:42.9 & 0.0051 & $12.8 \pm 2.6$ & $0.17$ & $0.23 \pm 0.09$ & 0.10  & 70 \\ 
\object{NGC 1792} & 05:05:14.45 & -37:58:50.7 & 0.0040 & $12.5 \pm 2.5$ & $0.17$ & $0.21 \pm 0.09$ & 0.12  & 63 \\ 
\object{NGC 4501} & 12:31:59.22 & +14:25:13.5 & 0.0076 & $15.3 \pm 2.8$ & $0.17$ & $0.21 \pm 0.08$ & 0.15  & 63 \\ 
\object{NGC 4536} & 12:34:27.13 & +02:11:16.4 & 0.0060 & $14.9 \pm 3.0$ & $0.15$ & $0.21 \pm 0.08$ & 0.13  & 73 \\ 
\object{NGC 5643} & 14:32:40.78 & -44:10:28.6 & 0.0040 & $14.4 \pm 1.0$ & $0.12$ & $0.19 \pm 0.03$ & 0.20  & 30 \\
\object{NGC 3628} & 11:20:17.02 & +13:35:22.2 & 0.0028 & $10.0 \pm 1.1$ & $0.14$ & $0.17 \pm 0.04$ & 0.09  & 79 \\ 
\object{NGC 1559} & 04:17:35.78 & -62:47:01.3 & 0.0043 & $12.7 \pm 2.5$ & $0.13$ & $0.16 \pm 0.07$ & 0.10  & 60 \\ 
\object{NGC 5248} & 13:37:32.07 & +08:53:06.2 & 0.0038 & $13.8 \pm 2.8$ & $0.12$ & $0.16 \pm 0.06$ & 0.14  & 56 \\ 
\object{NGC 3810} & 11:40:58.74 & +11:28:16.1 & 0.0033 & $15.4 \pm 3.1$ & $0.10$ & $0.13 \pm 0.05$ & 0.12  & 48 \\ 
\object{NGC 4654} & 12:43:56.64 & +13:07:34.9 & 0.0035 & $15.3 \pm 2.8$ & $0.10$ & $0.12 \pm 0.05$ & 0.13  & 60 \\ 
\object{NGC 1055}& 02:41:45.18 & +00:26:38.1 & 0.0033 & $11.3 \pm 2.3$ & $0.10$ & $0.12 \pm 0.05$ & 0.12  & 63 \\ 
\hline
\end{tabular}
  \tablefoot{
    \tablefoottext{a}{Luminosity distances calculated from the redshifts following the procedure outlined by \citet{san03}. For consistency, in galaxies where \citet{san03} instead use direct primary or secondary distance measurements, we use the same distances as they do. For the same reason, a $H_{0} =75$ km\,s$^{-1}$\,Mpc$^{-1}$, $\Omega_{\mathrm{m}} = 0.3$, and $\Omega_{\Lambda} = 0.7$ cosmology is adopted.}
    \tablefoottext{b}{Far-infrared luminosities used for selection, taken from \citet{san03}.}
    \tablefoottext{c}{Infrared luminosities calculated using the prescription in Table 1 of \citet[originally from \citealt{per87}]{san96} using IRAS fluxes taken from \citet{san03}.}
    \tablefoottext{d}{Ratio of IRAS fluxes at $25$ and $60$~$\mu$m taken from \citet{san03}.}
    \tablefoottext{e}{Inclinations of the host galaxies taken from the HyperLEDA database\footnote{http://leda.univ-lyon1.fr/} \citep{mak14}.}
    \tablefoottext{f}{NGC 1068 is a LIRG, but as we selected our samples based on $L_{\mathrm{FIR}}$, which is lower, it is included in the sub-LIRG sample. This is the only border case.}
    }
\end{center}
\label{tab:sample}
\end{table*}

\subsection{Selection bias}\label{sec:bias}
\begin{figure}
   \centering
   \includegraphics[width=\hsize]{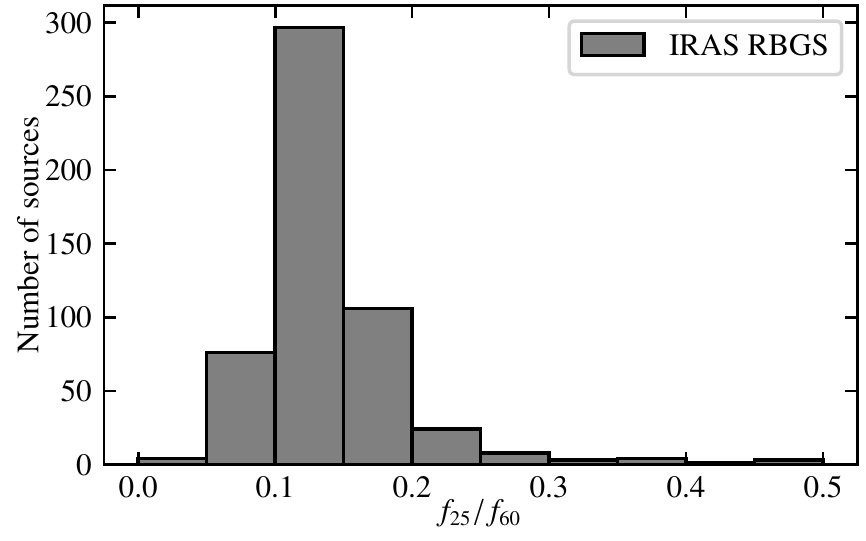}\\
   \includegraphics[width=\hsize]{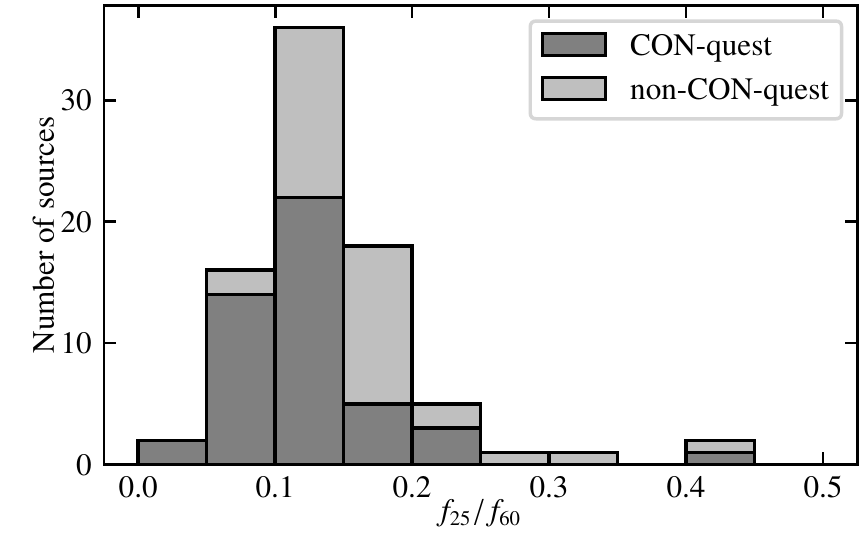}\\
   \includegraphics[width=\hsize]{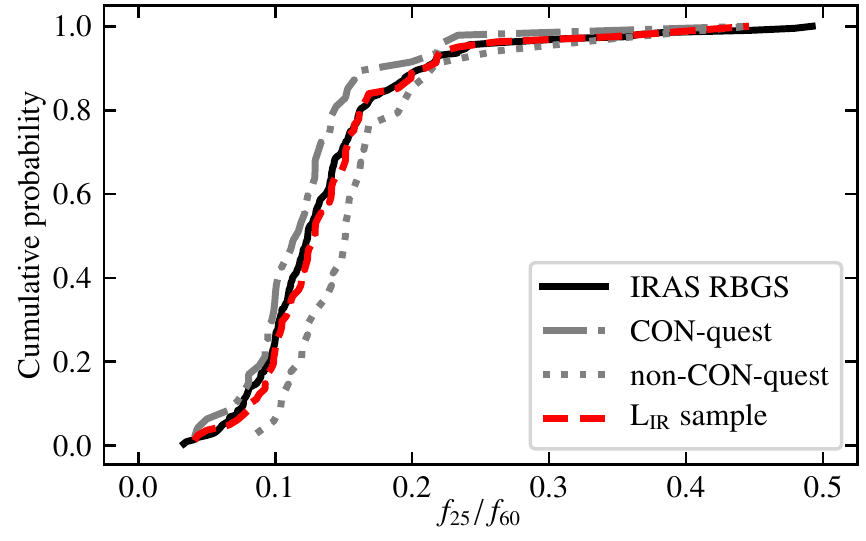}
   \caption{Distributions of IRAS $25/60$~$\mu$m ratios in various samples. \emph{Top:} histogram showing the distribution in the IRAS revised bright galaxy sample (RBGS). \emph{Middle:} stacked histogram showing the distribution in the sample that would have resulted if we had selected sources based on the total infrared luminosity (between $8$ and $1000$~$\mu$m) instead of the far-infrared luminosity (between $40$ and $400$~$\mu$m), with the CON-quest sample marked using a darker shade. \emph{Bottom:} empirical distribution functions of the RBGS (solid black), the CON-quest sample (dash-dotted gray), the additional sources required to complete an infrared selected sample (dotted gray), and the total infrared selected sample (dashed red). }
         \label{fig:bias}
\end{figure}

The selection based on far-infrared luminosity (between $40$ and $400$~$\mu$m) instead of total infrared luminosity (between $8$ and $1000$~$\mu$m) may introduce a bias against warm sources that emit a larger fraction of their radiation at shorter wavelengths. One way to investigate this potential infrared emission bias is to compare the distributions of the IRAS $25$ to $60$~$\mu$m flux density ratios ($f_{25}/f_{60}$) in our sample and the IRAS RBGS from which it was selected. In Fig. \ref{fig:bias} we make this comparison: the top panel shows the distribution in the RBGS, and the middle panel shows the distribution for the CON-quest sample using a dark shade. The $f_{25}/f_{60}$ distribution of the CON-quest sample is clearly skewed toward lower values when compared to the RBGS. In the middle panel we have also included the sources required, together with the CON-quest sample, to complete a sample selected based on the total (instead of far-infrared) infrared luminosities. The distribution of this combined sample is more similar to that of the underlying RBGS, as can also be seen in the bottom panel where we show the empirical distribution functions of the different samples.

\section{Observations and data reduction}\label{sec:observations}
For this survey, we targeted the HCN-vib $l=1f$ $J=3\text{--}2$ transition with rest frequency $267.199$~GHz. In the literature and the Atacama Large Millimeter/submillimeter Array (ALMA) archives, we found pre-existing observations of this transition in four ULIRGs, six LIRGs, and one sub-LIRG. Three sub-LIRGs had ALMA observations of the corresponding $J=4\text{--}3$ transition, and as the line was not detected in any of the sources we did not reobserve them in the $J=3\text{--}2$ transition. We have conducted new observations of 4 ULIRGs, 13 LIRGs, and 15 sub-LIRGs. The setup and the data reduction process of the new observations as well as our treatment of existing data are described in the following sections. 

\subsection{New ALMA observations}
New observations of the HCN-vib $l$\,=\,1$f$ $J$\,=\,3$-$2 transitions were performed as part of two ALMA band 6 projects, 2017.1.00759.S and 2018.1.01344.S. In total, 32 sources that fulfill the selection criteria of Sect. \ref{sec:sample} were observed in these two projects: four ULIRGs, 13 LIRGs and 15 sub-LIRGs:

\textit{2017.1.00759.S:} The four sources observed as part of this project were the ULIRGs IRAS~09022-3615, IRAS~13120-5453, IRAS~F14378-3651 and IRAS~17208-0014. The observations took place in September 2018 with the intermediate array configuration C43-5 (baseline lengths between $15$~m and $2$~km). The maximum recoverable scale (MRS, defined as $0.6\lambda/B_{\mathrm{min}}$, where $B_{\mathrm{min}}$ is the minimum baseline length) is $\sim 9$\arcsec\ ($6-13$~kpc).

\textit{2018.1.01344.S:} Thirteen LIRGs (ESO~173-G015, ESO~221-IG10, ESO~286-G035, ESO~320-G030, IC~4734, IC~5179, IRAS~F17138-1017, IRAS~17578-0400, NGC~2369, NGC~3110, NGC~5135, NGC~5734, UGC~2982) and 15 sub-LIRGs (NGC~660, NGC~1055, NGC~1559, NGC~1792, NGC~3627, NGC~3628, NGC~3810, NGC~4254, NGC~4303, NGC~4501, NGC~4527, NGC~4536, NGC~4654, NGC~4666, NGC~5248) were observed for this project between October 2018 and October 2019. The observations were carried out with the intermediate array configurations C43-4 and C43-5 with baseline lengths between $15$~m and $2.5$~km, resulting in an MRS of $\sim9$\arcsec\ ($0.4-3.5$~kpc).

The spectral setup for all 32 sources was the same: two spectral windows centered at the HCO$^{\rm +}$ ($\nu_{\mathrm{rest}}$\,=\,267.558~GHz) and HCN\,3$-$2 ($\nu_{\mathrm{rest}}$\,=\,265.886~GHz) transitions in the upper sideband. Two more spectral windows were used to recover the band 6 continuum emission in the lower sideband. The spectral windows were set with a bandwidth of $1.875$~GHz, and a channel width of $3.9$~MHz. During the observations, bright quasars were used for bandpass and flux calibration; weaker quasars close to each target source were used as phase/gain calibrators.

The imaging of the visibility sets for all 32 sources was performed using the ''tclean'' task in the Common Astronomy Software Applications \citep[CASA\footnote{http://casa.nrao.edu/};][]{mcm07} package version 5.4.0. Applying a natural weighting resulted in data cubes with beam sizes of roughly 0.3\arcsec\ (spatial scales between $\sim$45 and $\sim$95~pc) for the ULIRGs in 2017.1.00759.S, and beam sizes between $\sim$0.2\arcsec\ and $\sim$0.8\arcsec\ (spatial scales between $\sim$5 and $\sim$35~pc) for the LIRGs and sub-LIRGs in 2018.1.01344.S (see Table~\ref{tab:observations} for more information). For analysis purposes we Hanning-smoothed the data cubes to a velocity resolution of 20~km\,s$^{\rm -1}$. The resulting $1$-$\sigma$ noise levels per channel for each source are listed in Table~\ref{tab:observations}. After calibration and imaging within CASA, all image cubes were converted into FITS format for further analysis.

To obtain the HCN-vib fluxes and upper limits in sources with relatively narrow well separated spectral lines, we integrated the data cubes over the velocity range $\pm150$~km\,s$^{-1}$ around the systemic velocity of the transition, calculated from the redshifts given in Table \ref{tab:sample} in the heliocentric frame. This velocity range was set to avoid blending issues with the HCO$^{+}$ transition that is blueward of the HCN-vib line by approximately $400$~km\,s$^{-1}$. We then extracted the fluxes from a region with a diameter of twice the Gaussian full width at half maximum of the velocity integrated HCO$^{+}$ emission. To compensate for line blending in sources with strong and broad lines, we instead obtained the flux through a fit to the spatially integrated spectrum extracted from the same region. In the case of a nondetection of the HCN-vib line, we set a $3$-$\sigma$ upper limit to the flux based on the rms in the velocity integrated map and the line width of the ground state transitions.

We obtained estimates for the continuum properties from two-dimensional Gaussian fits to the continuum maps using the CASA task IMFIT which deconvolves the synthesized beam from the fitted component size, see \citet{con97} for a discussion of the error estimates in this process. In sources with multiple continuum components, we use the results from the strongest one, as this is where a potential CON would be most likely to be found.
  
\begin{table}[htbp!]
\caption{Properties of the new ALMA observations.}
\begin{center}
\begin{tabular}{lcc}
  \hline
  \hline
  \noalign{\smallskip}
Name & Beam size\tablefootmark{a} & Sensitivity\tablefootmark{a,b} \\ 
& (\arcsec) & (mJy\,beam$^{-1}$) \\
\hline
\noalign{\smallskip}
\multicolumn{3}{c}{ULIRGs} \\
\hline
\noalign{\smallskip}
IRAS 17208-0014 & $0.36 \times 0.33$ & $0.74$ \\ 
IRAS 13120-5453 & $0.36 \times 0.33$ & $0.76$ \\ 
IRAS 09022-3615 & $0.34 \times 0.29$ & $0.15$ \\ 
IRAS F14378-3651 & $0.31 \times 0.27$ & $0.19$ \\ 
\hline
\noalign{\smallskip}
\multicolumn{3}{c}{LIRGs} \\
\hline
\noalign{\smallskip}
IRAS F17138-1017 & $0.40 \times 0.31$ & $0.61$ \\ 
IRAS 17578-0400 & $0.37 \times 0.30$ & $0.48$ \\
ESO 173-G015 & $0.33 \times 0.30$ & $0.42$ \\ 
NGC 3110 & $0.41 \times 0.28$ & $0.40$ \\ 
IC 4734 & $0.40 \times 0.30$ & $0.53$ \\ 
NGC 5135 & $0.34 \times 0.28$ & $0.65$ \\ 
ESO 221-IG10 & $0.35 \times 0.32$ & $0.58$ \\ 
IC 5179 & $0.28 \times 0.24$ & $0.41$ \\ 
UGC 2982 & $0.37 \times 0.31$ & $0.33$ \\ 
ESO 286-G035 & $0.31 \times 0.27$ & $0.43$ \\ 
NGC 2369 & $0.28 \times 0.26$ & $0.52$ \\ 
NGC 5734 & $0.31 \times 0.29$ & $0.44$ \\ 
ESO 320-G030 & $0.35 \times 0.30$ & $0.54$ \\
\hline
\noalign{\smallskip}
\multicolumn{3}{c}{sub-LIRGs} \\
\hline
\noalign{\smallskip}
NGC 3628 & $0.61 \times 0.52$ & $0.51$ \\ 
NGC 4254 & $0.68 \times 0.55$ & $0.37$ \\ 
NGC 4303 & $0.63 \times 0.52$ & $0.35$ \\ 
NGC 660 & $0.78 \times 0.64$ & $0.47$ \\ 
NGC 4527 & $0.63 \times 0.52$ & $0.58$ \\ 
NGC 3627 & $0.70 \times 0.53$ & $0.51$ \\ 
NGC 4666 & $0.61 \times 0.54$ & $0.56$ \\ 
NGC 1792 & $0.55 \times 0.52$ & $0.47$ \\ 
NGC 4501 & $0.65 \times 0.57$ & $0.39$ \\ 
NGC 4536 & $0.60 \times 0.52$ & $0.60$ \\ 
NGC 1559 & $0.58 \times 0.44$ & $0.65$ \\ 
NGC 5248 & $0.70 \times 0.54$ & $0.60$ \\ 
NGC 3810 & $0.64 \times 0.55$ & $0.46$ \\ 
NGC 4654 & $0.64 \times 0.54$ & $0.41$ \\ 
NGC 1055 & $0.56 \times 0.49$ & $0.66$ \\
\hline
\end{tabular}
  \tablefoot{
    \tablefoottext{a}{With natural weighting of the interferometric visibilities.}
    \tablefoottext{b}{Given as $1$-$\sigma$ rms for channel widths of $20$~km\,s$^{-1}$.}
    }
\end{center}
\label{tab:observations}
\end{table}

\subsection{Archival and literature data}
Details about the observation setups and data reduction for sources with pre-existing data can be found in the references listed in the last column of Table \ref{tab:results}. For previous ALMA observations, we have not used the published values for the HCN-vib line fluxes in cases where these exist. We have instead extracted these in the same way as for newly observed sources, using the calibrated data sets obtained from the principal investigators of the projects in question. Some differences in the instrumental setup compared to our new observations exist, but none that are large enough to impact the appropriate extraction procedure.

\section{Results}\label{sec:results}
Luminosities of HCN-vib, continuum properties, and HCN-vib surface brightnesses are presented in Table \ref{tab:results}. For comparison with earlier work, the $L_{\mathrm{HCN-vib}}/L_{\mathrm{IR}}$ ratio is also listed. Line luminosities are calculated using Eq.\,1 of \citet{sol05}, applied to HCN-vib:
\begin{equation}
L_{\mathrm{HCN-vib}}=1.04\times10^{-3}S_{\mathrm{HCN-vib}}\,\Delta v\,\nu_{\mathrm{rest}}(1+z)^{-1}D_{\mathrm{L}}^{2},
\end{equation}
where $L_{\mathrm{HCN-vib}}$ is the HCN-vib luminosity measured in $L_{\odot}$, $S_{\mathrm{HCN-vib}}\,\Delta v$ is the velocity integrated flux in Jy\,km\,s$^{-1}$, $\nu_{\mathrm{rest}}$ is the rest frequency in GHz, and $D_{\mathrm{L}}$ is the luminosity distance in Mpc.

\begin{figure*}
   \centering
   \includegraphics[width=8.8cm]{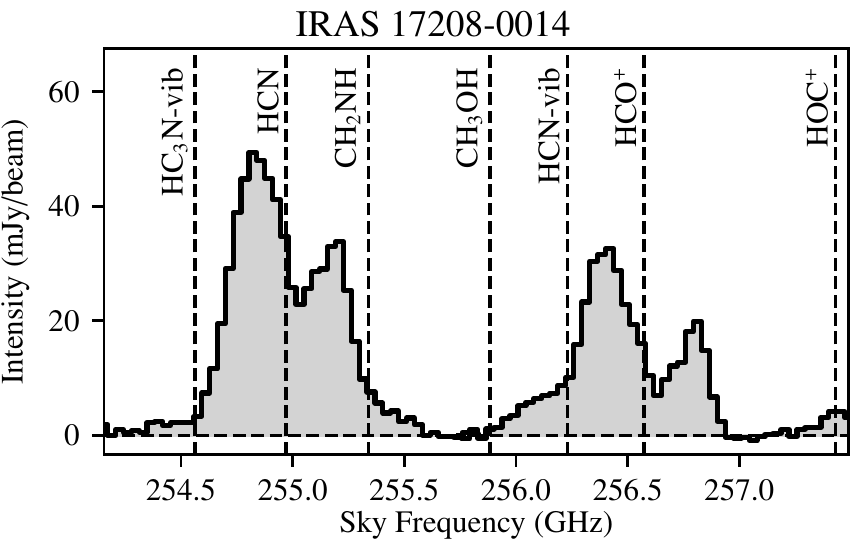}
   \includegraphics[width=8.8cm]{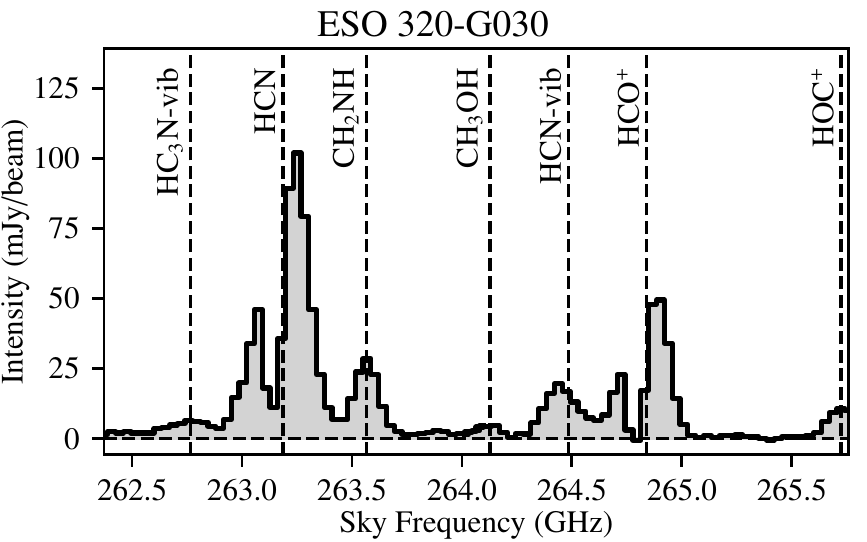}\\
   \includegraphics[width=8.8cm]{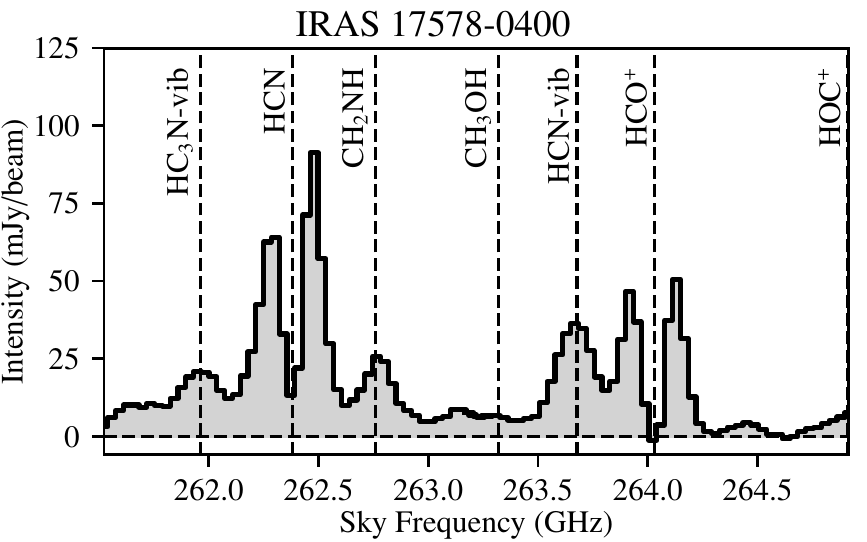}
   \includegraphics[width=8.8cm]{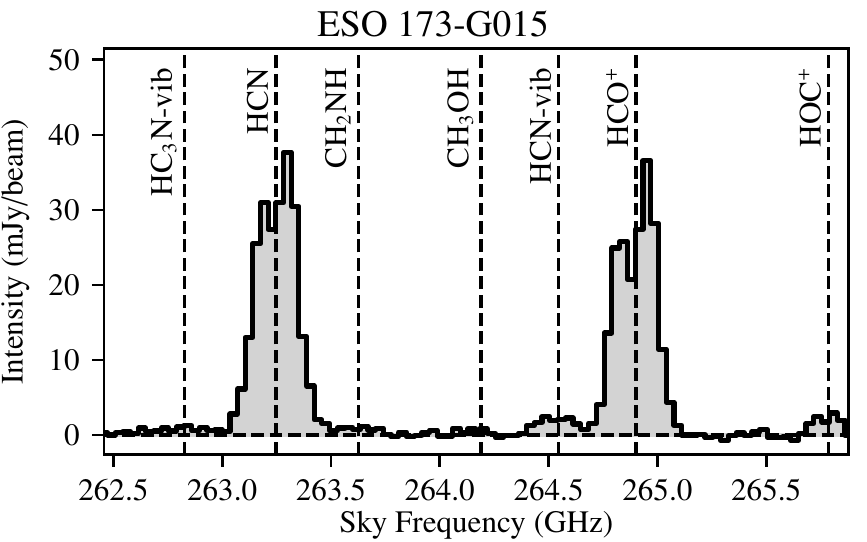}
   \includegraphics[width=8.8cm]{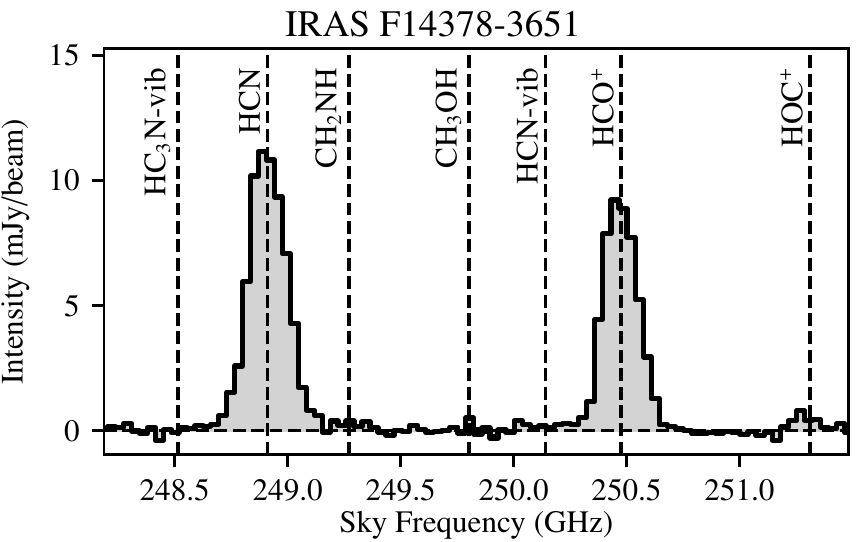}

   \caption{Continuum subtracted central spectra of galaxies with new HCN-vib $J=3\text{--}2$ detections. Frequencies of some lines commonly seen in CONs are marked with vertical dashed lines. In many of the sources, the ground state HCN and HCO$^{+}$ transitions are self-absorbed.}
         \label{fig:detections}
\end{figure*}

\begin{table*}[htbp!]
\caption{Results of HCN-vib $J=3\text{--}2$ and continuum measurements}
\begin{center}
\begin{tabular}{lcccccc}
  \hline
  \hline
  \noalign{\smallskip}
Name & $L_{\mathrm{HCN-vib}}$ & $S_{\mathrm{cont.}}$ & Continuum size & $\Sigma_{\mathrm{HCN-vib}}$ & $L_{\mathrm{HCN-vib}}/L_{\mathrm{IR}}$ & Ref. \\ 
& ($10^{3}$~L$_{\sun}$) & (mJy) & (mas $\times$ mas)  & (L$_{\sun}$\,pc$^{-2}$) & ($10^{-8}$) & \\
\hline
\noalign{\smallskip} 
\multicolumn{7}{c}{ULIRGs} \\
\hline
\noalign{\smallskip}
IRAS 17208-0014 & $62.2 \pm 11.0$ & $37.9 \pm 0.4$ & $273 \pm 6 \times 171 \pm6$ & $2.15\pm0.38$ & $2.5 \pm 0.3$ & 1 \\ 
IRAS F14348-1447 NE & $<5.776$ & $1.78 \pm 0.21$ &  $145 \pm29 \times 118 \pm 41$ &  $<0.19$ &$<0.283$ & 2 \\
IRAS F14348-1447 SW & $<9.17$ & $ 2.36 \pm 0.18 $ &  $113 \pm 20  \times 89 \pm 21 $ & $<0.37$ & $<0.451$ & 2 \\
IRAS F12112+0305 NE & $22.1 \pm 9.2$ & $9.29 \pm0.24$ & $201\pm 38 \times 261\pm41 $ & $0.23 \pm 0.11$ & $1.2 \pm 0.5$ & 3\\
IRAS F12112+0305 SW & $<9.071$ & $1.12 \pm 0.22$ & $ 445 \pm 212  \times 306 \pm 206 $ & $<0.04$  &$<0.481$ & 3\\
IRAS 13120-5453 & $<1.99$ & $32.62 \pm 0.88$ & $869 \pm 26 \times 749 \pm  22$ & $<0.01$ & $<0.108$ & 1 \\ 
IRAS 09022-3615 & $<5.552$ &  $5.69 \pm 0.16$ & $601 \pm 19 \times 312 \pm 15 $ & $<0.03$ & $<0.308$ & 1 \\ 
Arp 220 W& $61.7 \pm 8.4$ & $189.3\pm1.5$ & $220 \pm 23 \times 190\pm17$ & $11.96\pm 2.31$ & $3.8 \pm 0.1$ & 4 \\
Arp 220 E& $10.1 \pm 1.6$ & $80.5 \pm 1.7$ & $380 \pm 44 \times 230 \pm 30$ & $ 0.95 \pm 0.22$ & $0.63 \pm 0.06$ & 4 \\
IRAS F14378-3651 E& $3.9 \pm 0.6$ & $3.61 \pm 0.08$ & $350 \pm 16 \times 349 \pm17$ & $ 0.019\pm0.003$ & $0.27 \pm 0.05$ & 1\\
IRAS F14378-3651 W& $<1.1586$ &  $0.59 \pm 0.04$ & $345 \pm 36  \times 245 \pm 36$ & $<0.01$ & $<0.082$ & 1\\
IRAS F22491-1808 \tablefootmark{a}& $27.2 \pm 18.0$ & $4.41\pm0.07$ & $66 \pm5 \times 43\pm7$ & $4.84\pm 3.31$ & $2.1 \pm 1.3$  & 3\\
\hline
\noalign{\smallskip}
\multicolumn{7}{c}{LIRGs} \\
\hline
\noalign{\smallskip}
NGC 1614 & $<0.425$ & $16.6 \pm 1.4 $ & $ 1930 \pm 220  \times 1550 \pm 220 $ & $<0.002$ & $<0.105$ & 3 \\ 
NGC 7469 & $<0.121$ &$3.76 \pm 0.38$ & $342 \pm 143  \times 229\pm 197$ & $<0.02$ & $<0.031$ & 3 \\ 
NGC 3256 S & $<0.065$ &$9.28 \pm 0.79$ & $1247 \pm 123 \times 545\pm 69$ & $<0.004$ & $<0.017$ & 5\\
NGC 3256 N & $<0.065$ &$17.6 \pm 1.1$ & $1790 \pm 117 \times 1290\pm86 $ & $<0.001$ & $<0.017$ & 5\\
IRAS F17138-1017 & $<0.247$ & $3.75 \pm 0.4$ & $ 1217\pm131  \times 687\pm84 $ & $<0.003$ &$<0.095$ & 1\\
IRAS 17578-0400 & $15.2 \pm 2.1$ & $54.57\pm 0.71$ & $227\pm10 \times 212\pm 10$ & $4.76\pm0.72$ & $6.6 \pm 0.15$ & 1\\ 
NGC 7130 & $<0.203$ &$ 13.03\pm 0.77 $ & $745 \pm56  \times 583\pm45 $ & $<0.005$ & $<0.089$ & 6 \\ 
ESO 173-G015 & $1.2 \pm 0.2$ & $50.6\pm 1.5$ & $959\pm31 \times 333\pm13$ & $0.19\pm 0.03$ & $0.55 \pm 0.06$& 1 \\ 
NGC 3110 & $<0.157$ &$1.17 \pm 0.14 $ & $987 \pm132  \times 358\pm49 $ & $<0.004$ & $<0.077$ & 1 \\ 
IC 4734 & $<0.18$ & $10.64 \pm 0.84$ & $ 767\pm77  \times 621\pm69 $ & $<0.004$ &$<0.095$ & 1 \\ 
Zw 049.057 & $11.3 \pm 2.3$ & $37.2\pm 2.5$ & $294 \pm 83 \times 171\pm 61$ & $3.37 \pm 1.68$ & $6.0 \pm 0.9$ & 7 \\ 
NGC 5135 & $<0.126$ & $3.07 \pm0.42 $ & $669 \pm100  \times 368\pm78 $ & $<0.01$ &$<0.085$ & 1 \\ 
ESO 221-IG10 & $<0.143$ &$1.98 \pm 0.25$ & $ 1130\pm150  \times 472\pm69 $ & $<0.004$ & $<0.098$ & 1 \\ 
IC 5179 & $<0.055$ &$1.93 \pm 0.10$ & $408\pm23  \times 323\pm23 $ & $<0.01$ & $<0.038$ & 1 \\ 
UGC 2982 & $<0.124$ & $0.49 \pm0.07 $ & $800 \pm 130 \times 457\pm 89$ & $<0.003$ &$<0.088$ & 1 \\ 
ESO 286-G035 & $<0.15$ &$2.15 \pm 0.22$ & $1213 \pm 137  \times 304 \pm 45 $ & $<0.004$ & $<0.112$ & 1 \\ 
NGC 4418 \tablefootmark{a} & $4.9 \pm 1.1$ & $99.3\pm 0.5$ & $190\pm 9 \times 138\pm 9$ & $9.12\pm 2.23$ & $3.8 \pm 0.7$ & 8 \\ 
NGC 2369 & $<0.075$ &$11.82 \pm 1.0$ & $ 1387\pm122  \times 495\pm46 $ & $<0.003$ & $<0.06$ & 1 \\ 
NGC 5734 & $<0.115$ &$0.38 \pm 0.09 $ & $ 638\pm173  \times 166\pm84 $ & $<0.02$ & $<0.101$ & 1 \\ 
ESO 320-G030 & $2.0 \pm 0.3$ & $33.0\pm0.5$ & $282\pm7 \times 208\pm6$ & $1.4\pm0.2$ & $1.7 \pm 0.1$ & 1 \\
\hline
\noalign{\smallskip}
\multicolumn{7}{c}{sub-LIRGs} \\
\hline
\noalign{\smallskip}
NGC 1068 & $<0.01$ & $13.1 \pm 1.1$ & $172\pm24  \times 104\pm13 $ & $<0.13$ &$<0.004$ & 9 \\ 
NGC 1808\tablefootmark{b} & $<0.003$ & \ldots & \ldots & \ldots & $<0.005$ & 10,11 \\ 
NGC 4254 & $<0.005$ &$0.55 \pm 0.08 $ & $685 \pm 171  \times 421\pm224 $ & $<0.004$ & $<0.016$ & 1 \\ 
NGC 4303 & $<0.005$ & $1.03 \pm 0.08$ & $855 \pm 94  \times 555 \pm 78 $ & $<0.003$ &$<0.016$ & 1 \\ 
NGC 660 & $<0.004$ & $28.4 \pm 2.8 $ & $ 1594\pm 180  \times 786\pm 97 $ & $<0.001$ &$<0.014$ & 1 \\ 
NGC 4527 & $<0.009$ &$2.43 \pm 0.23$ & $1388 \pm149  \times 473\pm 84 $ & $<0.003$ & $<0.033$ & 1 \\ 
NGC 3627 & $<0.003$ &$7.15 \pm 0.24$ & $984 \pm 44 \times 587 \pm 39$ & $<0.003$ & $<0.014$ & 1 \\ 
NGC 613\tablefootmark{b} & $<0.004$ & \ldots & \ldots & \ldots &$<0.018$ & 10,12 \\ 
NGC 4666 & $<0.006$ & $1.1 \pm 0.11$ & $1383 \pm157  \times 552 \pm 95 $ & $<0.003$ &$<0.025$ & 1 \\ 
NGC 1792 & $<0.004$ & $1.33\pm0.09 $ & $704 \pm58  \times 507 \pm52 $ & $<0.004$ &$<0.021$ & 1 \\ 
NGC 4501 & $<0.006$ &$1.25 \pm 0.11 $ & $438\pm98  \times 258\pm134 $ & $<0.01$ & $<0.026$ & 1 \\ 
NGC 4536 & $<0.008$ & $4.84 \pm 0.31 $ & $2166 \pm145  \times 1301\pm90 $ & $<0.001$& $<0.039$ & 1 \\ 
NGC 5643\tablefootmark{b} & $<0.006$ &  \ldots & \ldots & \ldots &$<0.034$ & 13 \\ 
NGC 3628 & $<0.003$ & $38.3 \pm 1.5$ & $2470 \pm 106 \times 479\pm36 $ & $<0.001$ & $<0.017$ & 1 \\ 
NGC 1559 & $<0.006$ & $0.33 \pm0.02 $ & $793 \pm72  \times 375\pm53 $ & $<0.007$ & $<0.038$ & 1 \\ 
NGC 5248 & $<0.007$ & $4.07 \pm 0.29 $ & $1077 \pm 97  \times 619\pm 82 $ & $<0.003$ &$<0.045$ & 1 \\ 
NGC 3810 & $<0.007$ & $0.52 \pm 0.08 $ & $1380 \pm 24 \times 820\pm17 $ & $<0.001$ & $<0.050$ & 1 \\ 
NGC 4654 & $<0.006$ & $0.83 \pm 0.10 $ & $1230 \pm17  \times 690\pm13 $ & $<0.002$ &$<0.044$ & 1 \\ 
NGC 1055 & $<0.004$ & $3.8 \pm 0.28 $ & $994 \pm 85 \times 588\pm61 $ & $<0.003$ &$<0.034$ & 1 \\
\hline
\end{tabular}
\tablefoot{
    \tablefoottext{a}{Continuum properties from observations by \citet{ima18} and Sakamoto et al. (in prep.) }
    \tablefoottext{b}{Observed in the $J=4\text{--}3$ transition.} 
    }
\tablebib{
(1)~This work; (2) \citet{ima19}; (3) \citet{ima16a}; (4) \citet{mar16};
(5) \citet{har18}; (6) Privon, priv. comm.; (7) \citet{aal15b}; (8) \citet{sak10}; (9) \citet{ima16c}; (10) \citet{com19}; (11) \citet{aud20}; (12) \citet{aud19}; (13) Garc{\'\i}a-Burillo et al. in prep.
}
\end{center}
\label{tab:results}
\end{table*}
In total, emission from vibrationally excited HCN is detected in five of the eight ULIRGs, five of the $19$ LIRGs, and in none of the $19$ sub-LIRGs. The corresponding numbers when considering individual nuclei resolved by our observations are six of $12$ in ULIRGs and five of $20$ in the LIRGs. For three of the LIRGs, ESO 320-G030, ESO 173-G015, and IRAS 17578-0400, and one of the ULIRGs, IRAS F14378-3651, this is the first detection of vibrationally excited HCN. Another ULIRG, IRAS 17208-0014, has previously been detected in the HCN-vib $J=4\text{--}3$ transition, but this is the first detection of the $J=3\text{--}2$ transition. We also note that HCN-vib $J=3\text{--}2$ emission, with a luminosity of $2.2$~L$_{\sun}$, has recently been detected in NGC~1068 by \citet{ima20}. In Fig. \ref{fig:detections}, we present the spectra of sources with new detections of HCN-vib $J=3\text{--}2$ emission. The transition is clearly visible in the spectra of all sources except for IRAS~F14378-3651, in which the integrated intensity map was used to confirm the detection.

In IRAS~17208-0014 the line widths are large, causing noticeable blending of HCN-vib with the HCO$^{+}$ transition. Self- and continuum absorption by the latter \citep[see][for a more thorough discussion]{aal15b} complicates the situation further, warranting a short explanation of the features in this spectrum. The peaks on each side of the central frequency of the HCO$^{+}$ transition are both part of the HCO$^{+}$ feature. The peculiar shape is caused by an emission component, peaking close to the central frequency of the transition, combined with an absorption component that has its maximum at a slightly higher frequency. Finally, the prominent wing on the low-frequency side of the HCO$^{+}$ transition is HCN-vib emission.

\subsection{Detection rate of compact obscured nuclei}\label{sec:detection_rate}
Using the criterion $\Sigma_{\mathrm{HCN-vib}}> 1$~L$_{\sun}$\,pc$^{-2}$ in the $J=3\text{--}2$ transition, three ULIRGs, four LIRGs, and no sub-LIRGs from our sample are classified as CONs. This translates into CON detection rates of $38^{+18}_{-13}\%$ in the ULIRG sample, $21^{+12}_{-6}\%$ in the LIRG sample, and $0^{+9}_{-0}\%$ in the sub-LIRG sample. If we instead consider individual resolved nuclei, the detection rates are  $25^{+16}_{-8}\%$ in the ULIRG sample and $20^{+12}_{-6}\%$ in the LIRG sample. The $1$-$\sigma$ confidence intervals were estimated using the beta distribution quantile technique \citep{cam11}. For reference, with the old criterion, $L_{\mathrm{HCN-vib}}/L_{\mathrm{IR}}>10^{-8}$, the same seven CONs would have been identified, together with an additional one in IRAS F12112+0305~NE.

\subsection{Distributions of host galaxy properties}\label{sec:host_properties}
Armed with the CON detection rates, we can explore links with the properties of their host galaxies. As there were no detections in the sub-LIRG sample we only include the (U)LIRGs in this comparison.

\begin{figure}
   \centering
   \includegraphics[width=\hsize]{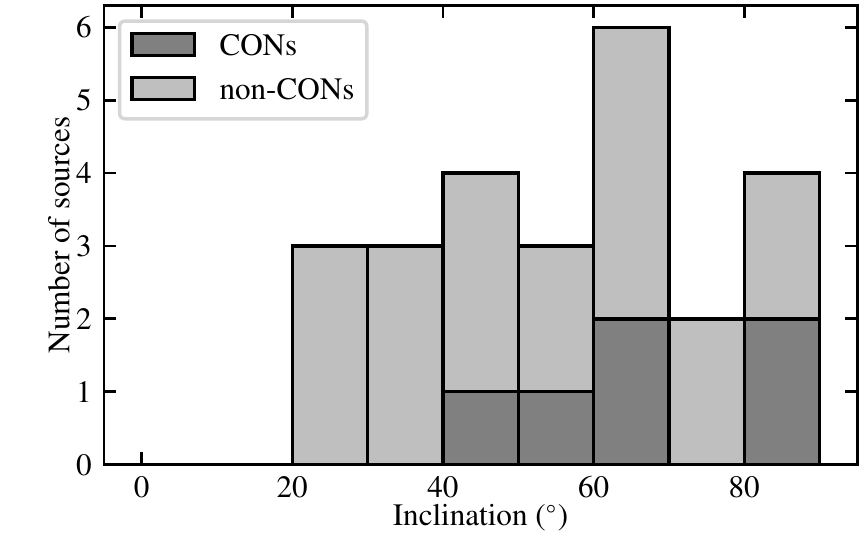}\\
   \includegraphics[width=\hsize]{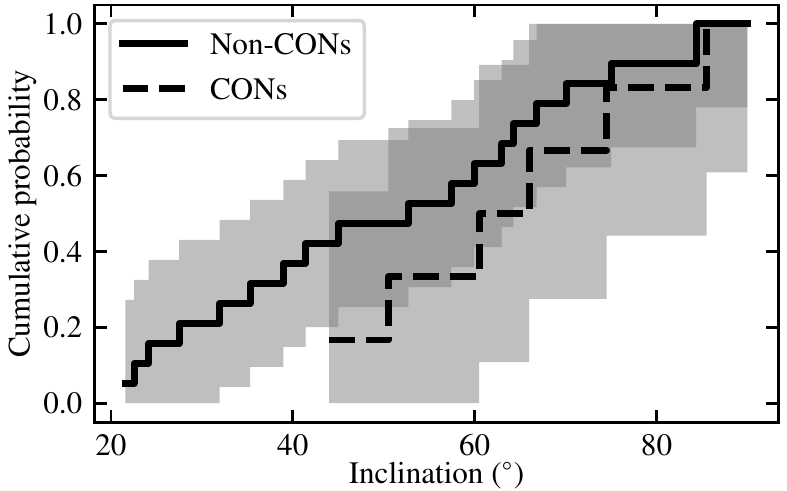}\\
   \caption{Distribution of inclinations in the (U)LIRG sample of CON-quest. \emph{Top:} stacked histogram of the CON-quest sample with CONs marked using a darker shade. \emph{Bottom:} empirical distribution functions of the CON and non-CON parts of the sample. The $1$-$\sigma$ confidence intervals are marked with a gray shade.}
         \label{fig:inclinations}
\end{figure}

\citet{fal19} suggested that galaxies with high inclination may be more likely to also have high $L_{\mathrm{HCN-vib}}/L_{\mathrm{IR}}$ ratios, and thus be classified as CONs. To assess whether CONs are preferentially found in high inclination systems, we can compare the distribution of inclinations for CONs with that of the rest of the sample galaxies. To do this, we use optical estimates of the inclinations taken from the HyperLEDA database\footnote{http://leda.univ-lyon1.fr/} \citep{mak14}. An important caveat is that inclinations are difficult to determine in the disturbed interacting galaxies that are common in the most luminous systems. The histogram and the empirical distribution functions, which are estimates of the true cumulative distribution functions, in Fig. \ref{fig:inclinations} suggest that, while high inclinations might be preferred, the distribution of inclinations is statistically not significantly different. An Anderson-Darling test \citep[e.g.,][]{sch87} applied to the data confirms this, giving a probability of more than $25\%$ that the populations are drawn from the same underlying distribution. However, we caution that the nuclear orientation may not be well connected to that of the larger scale structure \citep[e.g.,][]{pja17}.
 
\begin{figure}
   \centering
   \includegraphics[width=\hsize]{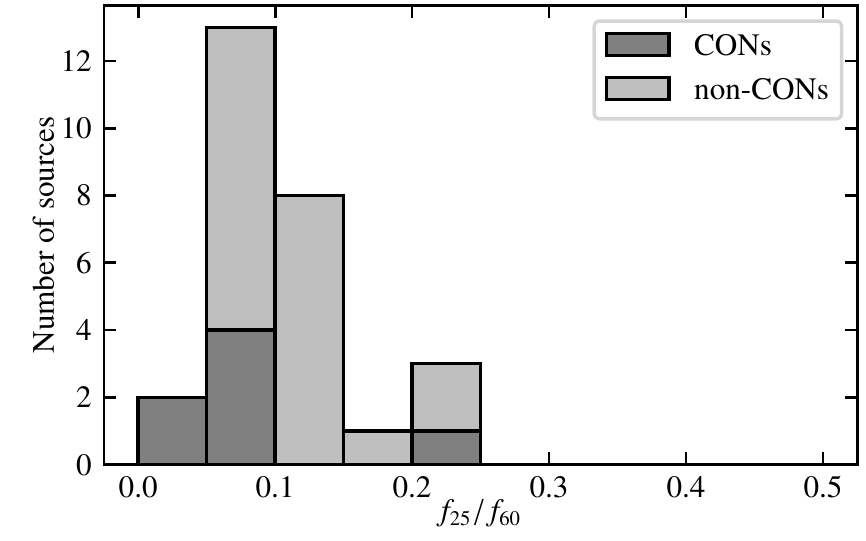}\\
   \includegraphics[width=\hsize]{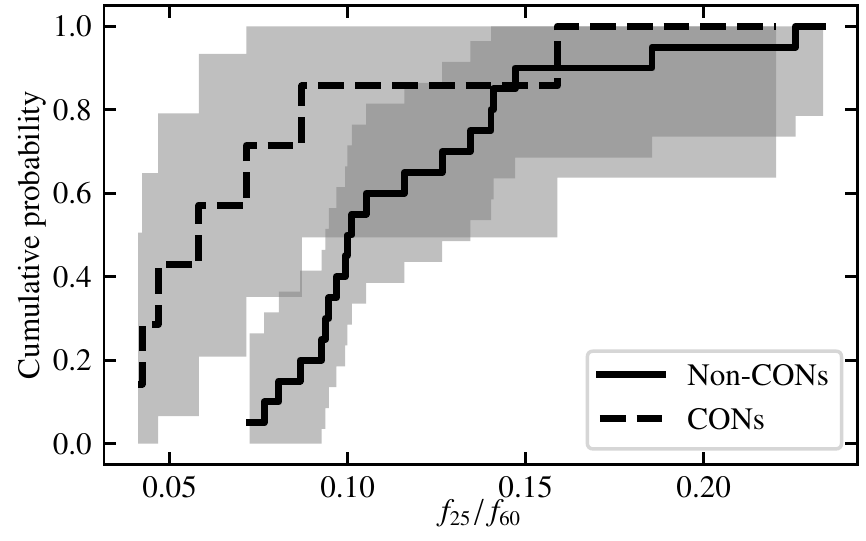}\\
   \caption{Distribution of IRAS $25/60$~$\mu$m ratios in the (U)LIRG sample of CON-quest. \emph{Top:} stacked histogram of the CON-quest sample with CONs marked using a darker shade. \emph{Bottom:} empirical distribution functions of the CON and non-CON parts of the sample. The $1$-$\sigma$ confidence intervals are marked with a gray shade.}
         \label{fig:iras_ratio}
\end{figure}

Another interesting property in the context of CONs, which are thought to contain warm cores with intense mid-infrared radiation fields, is the relative strength of the mid- and far-infrared emission as traced by the IRAS $25$ to $60$~$\mu$m flux density ratio ($f_{25}/f_{60}$). Visual inspection of both the histogram and the empirical distribution functions in Fig. \ref{fig:iras_ratio} suggests that the CONs and non-CONs have different distributions, with most of the CONs having lower $f_{25}/f_{60}$ ratios than the non-CONs. An Anderson-Darling test between the $f_{25}/f_{60}$ distributions of the CONs and the rest of the sample confirms this notion; the probability that the populations are drawn from the same underlying distribution is less than $1\%$. This result is interesting in the light of the sample being biased toward objects with lower $f_{25}/f_{60}$ ratios, see Sect. \ref{sec:effects_bias} for further discussion.

\begin{figure}
   \centering
   \includegraphics[width=\hsize]{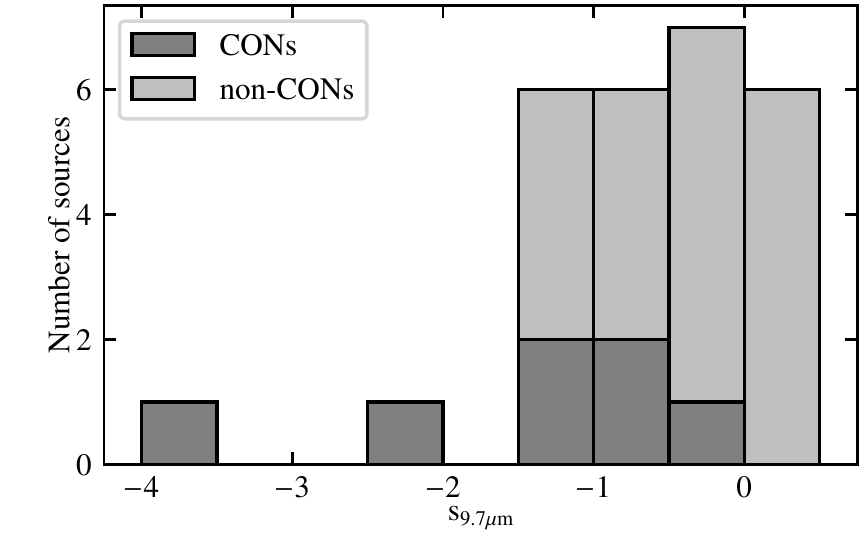}\\
   \includegraphics[width=\hsize]{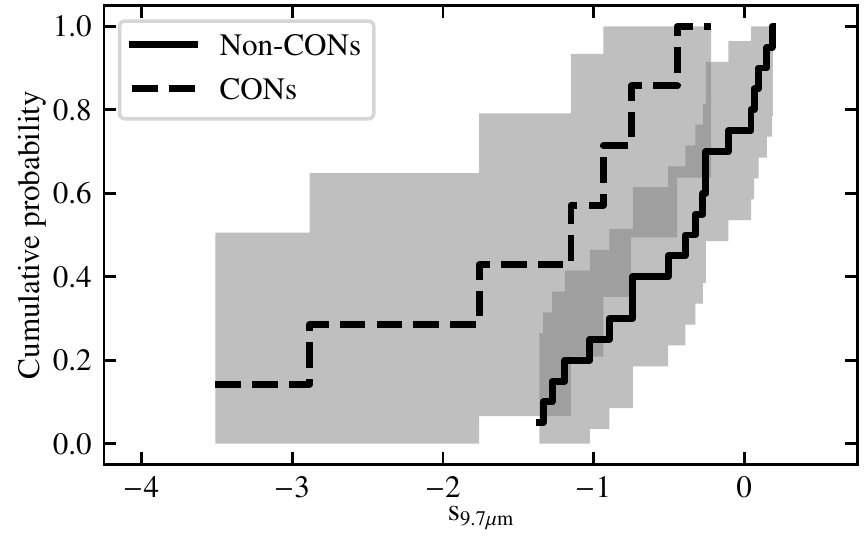}\\
   \caption{Distribution of the apparent depth of the $9.7$~$\mu$m silicate absorption feature in the (U)LIRG sample of CON-quest. \emph{Top:} stacked histogram of the CON-quest sample with CONs marked using a darker shade. \emph{Bottom:} empirical distribution functions of the CON and non-CON parts of the sample. The $1$-$\sigma$ confidence intervals are marked with a gray shade.}
         \label{fig:silicates}
\end{figure}

\begin{figure}
   \centering
   \includegraphics[width=\hsize]{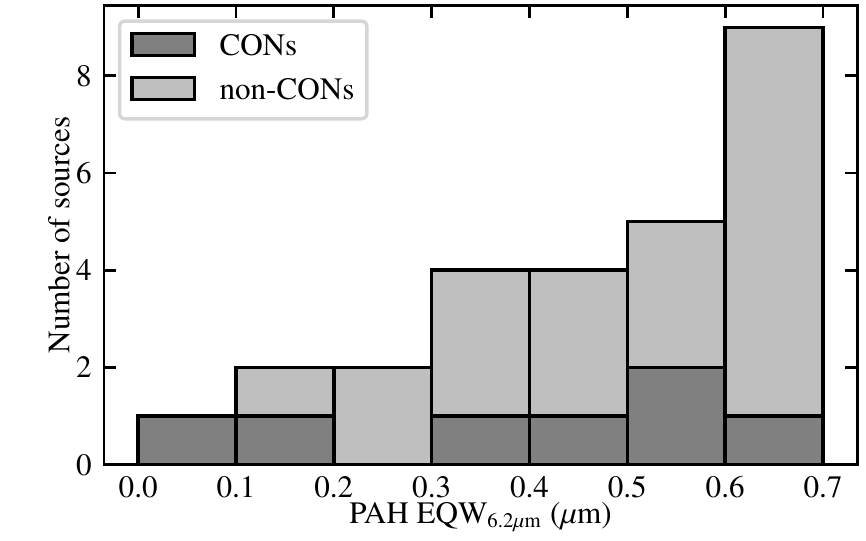}\\
   \includegraphics[width=\hsize]{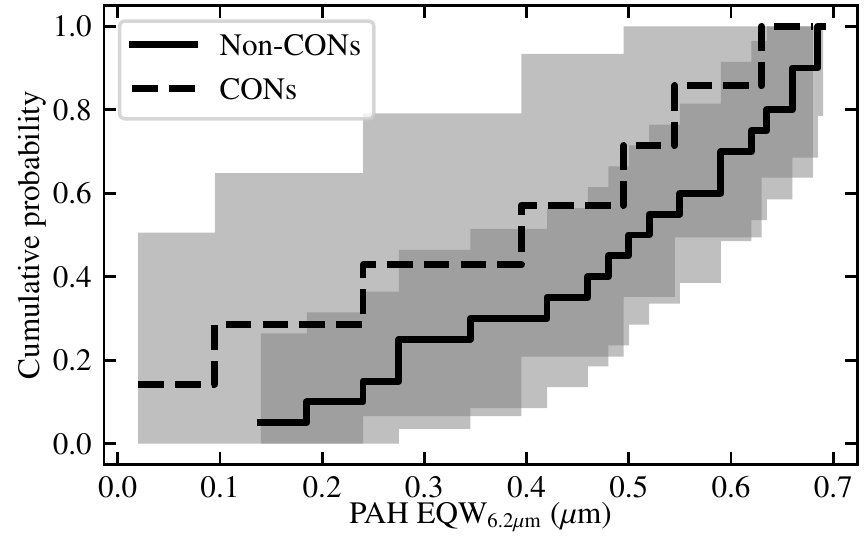}\\
   \caption{Distribution of the equivalent width of the $6.2$~$\mu$m PAH feature in the (U)LIRG sample of CON-quest. \emph{Top:} stacked histogram of the CON-quest sample with CONs marked using a darker shade. \emph{Bottom:} empirical distribution functions of the CON and non-CON parts of the sample. The $1$-$\sigma$ confidence intervals are marked with a gray shade.}
         \label{fig:pah}
\end{figure}

As the CONs are characterized by their high column densities of obscuring material, it may be of interest to investigate the strength of the $9.7$~$\mu$m silicate feature ($s_{9.7\mathrm{\mu m}}$) and the equivalent width of the $6.2$~$\mu$m polycyclic aromatic hydrocarbon (PAH) feature (EQW$_{6.2\mathrm{\mu m}}$), two properties used in the mid-infrared classification plot devised by \citet{spo07} to separate AGN-dominated, starburst-dominated, and deeply obscured nuclei. The strength of the silicate feature is defined as $s_{9.7\mathrm{\mu m}} = \ln{(f_{9.7\mathrm{\mu m}}/C_{9.7\mathrm{\mu m}})}$ where $f_{9.7\mathrm{\mu m}}$ is the measured flux density at $9.7$~$\mu$m and $C_{9.7\mathrm{\mu m}}$ is the continuum flux density at the same wavelength in the absence of an absorption feature. In the diagnostic plot of \citet[][their Fig. 1]{spo07}, deeply obscured nuclei are characterized by strong silicate absorptions and low PAH equivalent widths. In Figs. \ref{fig:silicates} and \ref{fig:pah} we present the histograms and empirical distribution functions of these two properties for the (U)LIRG samples, using data collected by \citet{sti13}. While the distributions of $s_{9.7\mathrm{\mu m}}$ seem to differ between the CONs and the rest of the sample, the distributions of EQW$_{6.2\mathrm{\mu m}}$ are more similar. Anderson-Darling tests indicate that the probabilities that the CON and non-CON populations are drawn from the same underlying distributions are lower than $5\%$ and higher than $25\%$, respectively, for $s_{9.7\mathrm{\mu m}}$ and EQW$_{6.2\mathrm{\mu m}}$.

\subsection{Noncircular motions in galaxies with compact obscured nuclei}\label{sec:noncircular}

An important way in which the CON and the extended host galaxy may affect each other is by in- or outflows of gas onto or from the nucleus. The presence of such noncircular motions could also provide clues to how the CONs may evolve with time. \citet{fal19} noted that galaxies with bright HCN-vib emission tend to show redshifted OH ground state absorption lines at $119$~$\mu$m, indicating possible molecular inflows. One of the CONs in our survey, IRAS 17578-0400, does not have any observations of this doublet. However, it has observations of the other ground state doublet at $79$~$\mu$m in the Herschel science archive (see appendix \ref{app:OH}), showing a small redshift with respect to the systemic velocity of the galaxy. 

\begin{figure}
  \centering
    \includegraphics[width=\hsize]{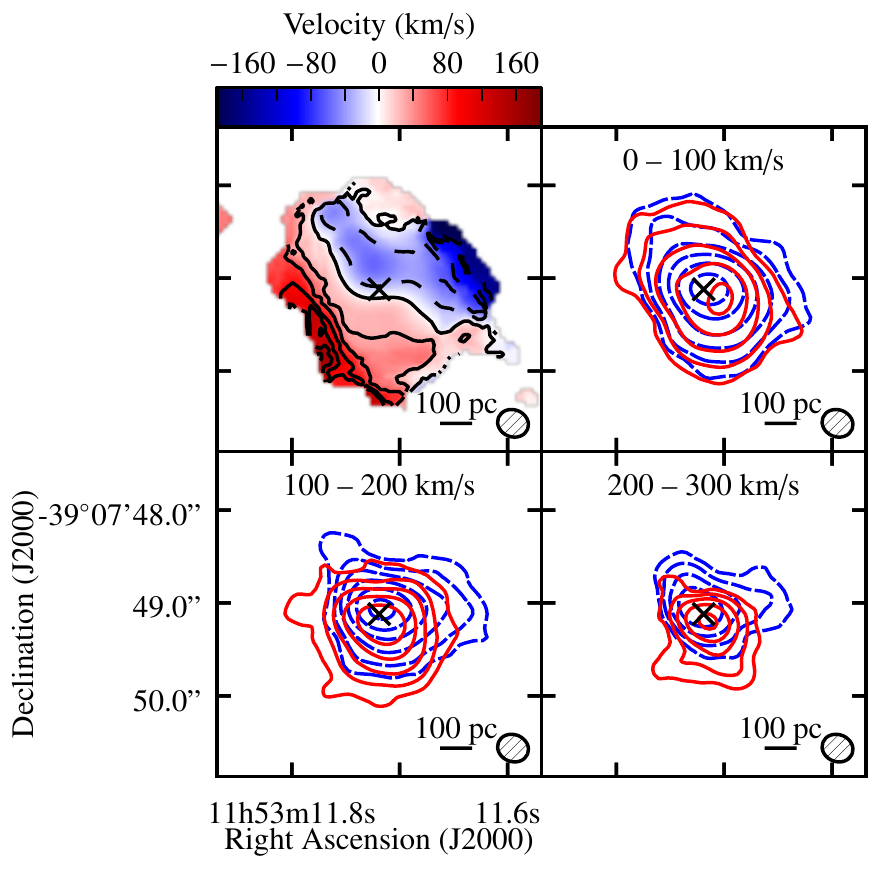}
  \caption{Intensity weighted velocity (moment 1) map and integrated intensity contours of HCN taken over the velocity ranges $0\text{--}100$, $100\text{--}200$, and $200\text{--}300$~km\,s$^{-1}$ from the systemic velocity (${\sim}3080$~km\,s$^{-1}$) on the red- (solid) and blueshifted (dashed) sides in ESO~320-G030. Velocity contours are given every $30$~km\,s$^{-1}$. Integrated intensity contours start at $0.24$, $0.16$, and $0.11$~Jy\,beam$^{-1}$\,km\,s$^{-1}$ ($5$-$\sigma$) for the three velocity intervals, respectively, and increase by factors of two.}
         \label{fig:outflow_ESO320}
\end{figure}

\begin{figure}
  \centering
  \includegraphics[width=\hsize]{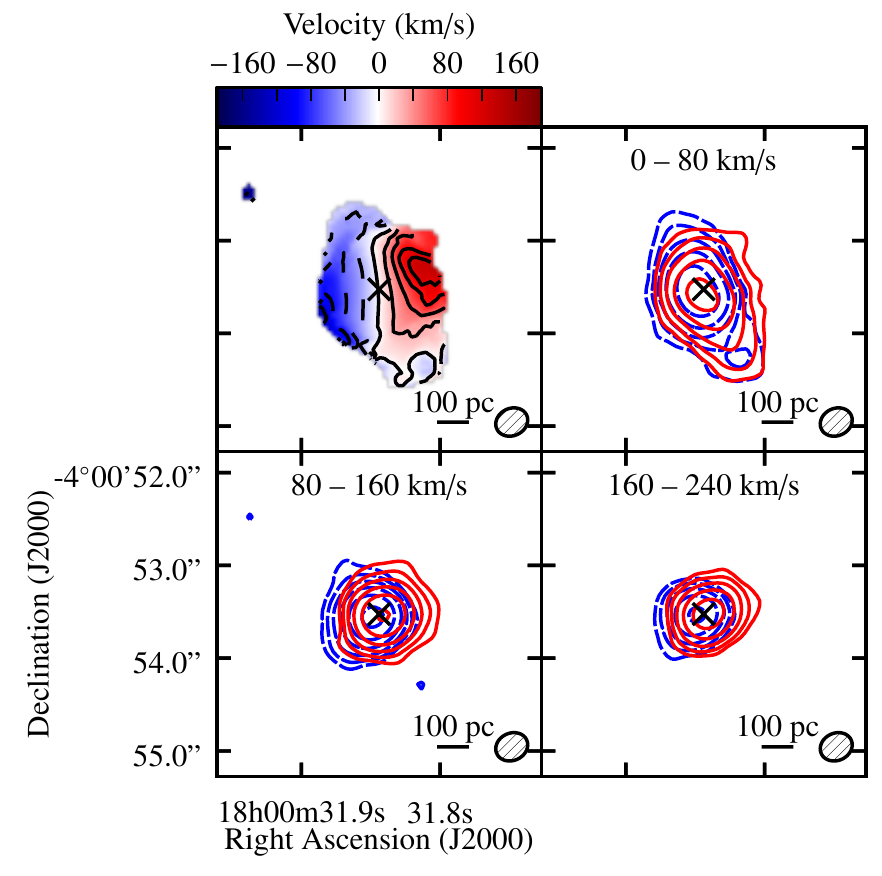}
  \caption{Intensity weighted velocity (moment 1) map and integrated intensity contours of HCN taken over the velocity ranges $0\text{--}80$, $80\text{--}160$, and $160\text{--}240$~km\,s$^{-1}$ from the systemic velocity (${\sim}4000$~km\,s$^{-1}$) on the red- (solid) and blueshifted (dashed) sides in IRAS~17578-0400. Velocity contours are given every $30$~km\,s$^{-1}$. Integrated intensity contours start at $0.16$, $0.20$, and $0.16$~Jy\,beam$^{-1}$\,km\,s$^{-1}$ ($5$-$\sigma$) for the three velocity intervals, respectively, and increase by factors of two.}
         \label{fig:outflow_I17578}
\end{figure}

Another result from \citet{fal19} is that at longer wavelengths CONs instead tend to show evidence of molecular outflows, often compact and collimated. The sensitivity of our new observations allows for the detection of faint spectral features, probing possible low mass outflows. For IRAS~17208-0014, this is hindered by the large line widths which cause severe blending with nearby lines of CH$_{2}$NH and vibrationally excited HC$_{3}$N, but a possible outflow signature in the form of blueshifted (${\sim}-70$~km\,s$^{-1}$) self-absorption features can be seen in the HCN and HCO$^{+}$ ground-state lines in Fig. \ref{fig:detections}. In ESO~320-G030 (Fig.\ref{fig:outflow_ESO320}) we see that the HCN emission at projected velocities between $0$ and $\pm100$~km\,s$^{-1}$ and between $\pm200$ and $\pm300$~km\,s$^{-1}$ from the systemic velocity has a component that is elongated along an axis with a position angle (PA) of $\sim 40\degr$ east of north. This is almost perpendicular to the major kinematic axis which has a PA of $\sim 135\degr$ as determined from the velocity field seen outside of the most nuclear region, consistent with the value found by \citet{per16} from observations of CO $J=2\text{--}1$. The orientation of the elongated structure is similar to that of the outflow reported by \citet{per16} which has a PA of between $10\degr$ and $40\degr$. Interestingly, the absorption features in the HCN and HCO$^{+}$ ground-state lines in ESO~320-G030 are redshifted with respect to the emission, possibly indicating that the absorbing gas is moving inward. We do not see any signatures of misaligned high-velocity emission in IRAS~17578-0400 but there is a tentative signature at lower velocities. In Fig. \ref{fig:outflow_I17578} we see that the emission within $\pm80$~km\,s$^{-1}$ (projected) of the systemic velocity is elongated along the kinematic minor axis (PA $\sim 25\degr$, again determined from the velocity field outside the most nuclear region), with a possible offset between the red- and blueshifted emission components along this axis. For this galaxy, we have not found any published observations of CO to compare the orientations of the kinematic axes in the nucleus and on larger scales. The absorption features in the HCN and HCO$^{+}$ ground-state lines of IRAS~17578-0400 do not show any strong velocity shifts, although the one in the HCN line appears slightly redshifted compared to the emission. 

\section{Discussion}\label{sec:discussion}
\subsection{How common are compact obscured nuclei?}
With the currently used definition (see Sect. \ref{sec:definition}), a remarkably high fraction of local LIRG and ULIRG systems, $\sim20\%$ and $\sim40\%$, respectively, are categorized as CONs. However, the (U)LIRGs themselves are relatively rare \citep[see, for example, the luminosity function in Fig. 12 by][]{san03}, and in lower luminosity galaxies the CON prevalence goes down significantly to a value close to zero. With a (U)LIRG number density of $\sim 6 \times10^{-5}$~Mpc$^{-3}$ in the local Universe \citep{san03}, we should thus expect a CON density of $\sim 10^{-5}$~Mpc$^{-3}$, taking into account the lower detection rate in less luminous systems. As discussed in Sect. \ref{sec:bias}, there is a potential bias due to our selection criteria which misses ``warm'' galaxies. If CONs are predominantly found in cold sources, as suggested in Sect. \ref{sec:where}, the inclusion of the warm sources will result in an overall decrease of $\sim50\%$ in the CON detection rate. On the other hand, it is possible that the only ``warm'' CON so far, NGC~4418, is not an exception but rather an example of a group of sources belonging to the second peak of a possible bimodal distribution. If this is the case, the local number density of CONs may even be slightly higher than estimated here.

Local CONs are important in order to conduct detailed studies, but in the context of galaxy evolution it is also important to determine how common they were at earlier times when the star formation rate and rate of black hole growth in the Universe was higher \citep[e.g.,][]{cha01,mar04}. At $z\sim1$, the number density of (U)LIRGs is up to two orders of magnitude larger than at $z<0.1$ \citep{lef05,mag09,mag13}. It is therefore likely that, if the CON fraction in (U)LIRGs remains high at high redshift, CONs may have played an important role in the evolution of galaxy nuclei. A point of caution, however, is that ULIRGs in the early Universe appear to have structures that are different from those of their local counterparts \citep[e.g.,][]{ruj11}.

\subsubsection{Sources just below the limit}
Our sample contains one LIRG: ESO~173-G015, and two ULIRG nuclei: Arp~220 E, and IRAS F12112+0305 NE, which have clear HCN-vib detections but with values of $\Sigma_{\mathrm{HCN-vib}}$ that are up to a factor of five too low to be classified as CONs. A question that should be asked is how these objects should be classified and what their relation to the CONs is. If we compare the surface brightnesses to those of the CONs, we see that the weakest CON has a $\Sigma_{\mathrm{HCN-vib}}$ value that is less than a factor of two larger than that of the strongest non-CON. Interestingly, when considering the $L_{\mathrm{HCN-vib}}/L_{\mathrm{IR}}$ ratio using the infrared emission from the individual nuclei, Arp~220 E has a higher ratio than the neighboring CON Arp~220 W \citep{mar16}. In addition to the HCN-vib emission, the spectra of these galaxies also exhibit some other features commonly seen in CONs. These include possible CH$_{2}$NH lines, double peaked profiles in the HCN and HCO$^{+}$ ground-state lines (see Fig. \ref{fig:detections}), and, at least in ESO~173-G015, a clearly detected HOC$^{+}$ line. With these similarities in mind, it is possible that these objects are related to the CONs, either as less extreme versions, as former CONs, or as objects in a transition phase to become CONs. With an empirically based, but still somewhat arbitrary, limit separating CONs from non-CONs, it may also be fruitful to think about the CONs as a population of sources at one of the extreme ends of a continuous distribution of source properties. The sources that fall just below the limit to be considered a CON would then not be qualitatively different from those that are above the same limit. For example, one possibility is that these objects are as obscured as the regular CONs, but without a central engine powerful enough to create the environment required for the HCN-vib emission to rise above the limit in our definition. In any case, further studies of these, and similar, objects may be important for our understanding of the timescales and processes of formation and destruction of CONs.

\subsection{Where do compact obscured nuclei occur?}\label{sec:where}
There is no statistically significant evidence that CONs are found primarily in highly inclined systems, which was one of the explanations for the lack of wide angle outflows in CONs suggested by \citet{fal19}. However, it is important to remember that it is difficult to determine the inclination of the disturbed interacting galaxies that are common in the (U)LIRG samples. These systems may also contain two nuclei with distinct orientations that will not necessarily be well-resolved. Furthermore, the obscuration in CONs typically occurs on scales of $\sim 100$~pc or less \citep[e.g.,][]{aal19}, and the orientation of this structure may differ from that of the host galaxy \citep[e.g.,][]{pja17}. This will be further investigated in a follow-up project with an angular resolution of ${\sim}0.02$\arcsec\ (${\sim}5$~pc), targeting the HCN-vib $J=3\text{--}2$ transition as well as the continuum at $1.1$ and $3$~mm.

When it comes to the relative strength of the mid- and far-infrared emission, there is strong evidence that CON galaxies have a different distribution of $f_{25}/f_{60}$ ratios compared to the rest of the sample galaxies. The majority of the detected CONs, as well as the other sources with detected HCN-vib emission, have mid to far-infrared continuum ratios smaller than $0.1$, indicating relatively cool dust emission. This result may seem somewhat surprising at first, given that efficient excitation of HCN-vib requires intense mid-infrared radiation. However, suppression of the mid-infrared continuum is a natural consequence of the high column densities that are found in CONs. Similar arguments were invoked already by \citet{bry99} to explain the low $f_{12}/f_{25}$ ratios in other LIRGs with high central gas surface densities. The low $f_{25}/f_{60}$ ratios are also consistent with the greenhouse scenario of \citet{gon19} where the mid- and far-infrared ``photospheres'' are significantly cooler than the interior regions. As they point out, however, their study assumes spherical symmetry and no clumpiness. In a more realistic situation, for example with a clumpy medium or a disk-like structure, some amount of mid-infrared radiation leaking out through sightlines with lower column density is expected. An indication that this is occuring comes from the results of \citet{lah07}, who find excitation temperatures of $200-300$~K in their analysis of the $14$~$\mu$m HCN absorption band, comparable to the central temperature found in \object{IC~860} by \citet{aal19} from millimeter observations. This leaking would reduce the greenhouse effect, but the radiative transfer in disk-outflow systems as the one suggested in IC~860 \citep{aal19} is complex, and requires further study. We note also that one of the CONs, NGC~4418, stands out with a substantially higher $f_{25}/f_{60}$ ratio of $0.22$. This could be due to a larger fraction of mid-infrared radiation leaking out in this source, either due to a lower column density or a more clumpy medium, or to radiation from surrounding structures as, for example, the super star clusters suggested by \citet{var14} based on radio VLBI observations. It could also be that the difference is due to different fractions of the total $L_{\mathrm{IR}}$ coming from the CON and the host galaxy. For example, we see in Fig. 2b of \citet{gon19} that, although lower than in less obscured nuclei, the intrinsic $f_{25}/f_{60}$ ratio of the model with high column density is still close to the $0.22$ observed in NGC~4418. It is thus possible that the CON in NGC~4418 is dominating the infrared luminosity of the galaxy, while the CONs in other galaxies are ``diluted'' by (colder) emission from the host. In any case, further studies are required to investigate if, and in that case how, cold and warm CONs differ.

\begin{figure}
  \centering
    \includegraphics[width=\hsize]{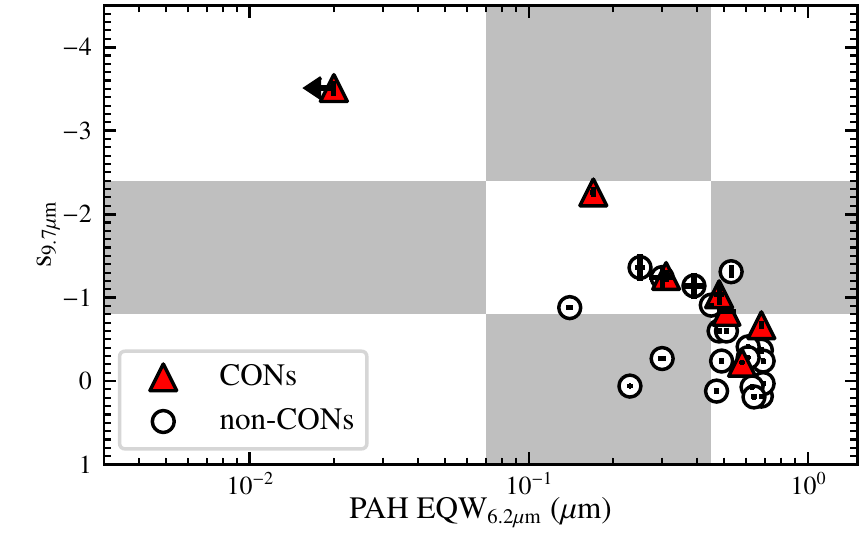}
  \caption{Diagnostic plot of the strength of the $9.7$~$\mu$m silicate feature against the equivalent width of the $6.2$~$\mu$m PAH feature suggested by \citet{spo07}. The CON-quest LIRGs and ULIRGs are plotted using data presented by \citet{sti13}, with CONs plotted as red triangles and non-CONs plotted as white circles. Shaded and white rectangles indicate the approximate regions of the plot used by \citet{spo07} to divide sources into different classes.}
         \label{fig:spoon_diagram}
\end{figure}

If we turn to the mid-infrared diagnostics used by \citet{spo07}, it seems that CONs generally have similar EQW$_{6.2\mathrm{\mu m}}$ but stronger $s_{9.7\mathrm{\mu m}}$ when compared to the rest of the sample. This may be an indication that the CONs are distributed along most of the diagonal branch, going from strong $s_{9.7\mathrm{\mu m}}$ and low EQW$_{6.2\mathrm{\mu m}}$ to weak $s_{9.7\mathrm{\mu m}}$ and high EQW$_{6.2\mathrm{\mu m}}$, in the diagnostic plot by \citet[][their Fig. 1]{spo07}. This notion is confirmed if we look at the combination of the two values for each individual CON (Fig. \ref{fig:spoon_diagram}). Interestingly, the galaxies along this diagonal branch are described by \citet{spo07} as ``intermediate stages between a fully obscured galactic nucleus and an unobscured nuclear starburst'', a statement that is seemingly inconsistent with the fact that the highly obscured CONs are found along most of the range of this branch. This could however, again, be explained by the greenhouse scenario in which most of the mid-infrared emission from CONs is hidden by the outer, cooler, parts of the nucleus. We note that NGC~4418 has the deepest silicate absorption of the CONs, possibly favoring the leakage explanation for its high $f_{25}/f_{60}$ ratio as the $9.7$~$\mu$m continuum against which the absorption takes place would be very low in a pure greenhouse scenario \citep[Fig. 2b of][]{gon19}.

There are also striking differences between the three luminosity bins: none of the sub-LIRGs, $\sim20\%$ of the LIRGs, and $\sim40\%$ of the ULIRGs in our sample host CONs. This may be related to the conditions necessary for CONs to form, and it is therefore worth examining what sets the three subsamples apart. Besides the trivial differences in luminosity, they mainly differ in the morphologies of the constituent galaxies. While most sub-LIRGs are single gas-rich spirals, the fraction of interacting galaxies, as well as the severity of the interaction, increases with luminosity in the LIRG luminosity range, to dominate the population of ULIRGs \citep{san96}. If the CON phenomenon is primarily linked to rapid gas inflows we expect to find them primarily in actively interacting systems, and this is consistent with the higher fraction seen in ULIRGs, which are all major mergers. It is also possible that the higher detection rate in ULIRGs is due to the larger fraction of systems with multiple nuclei, as the rate decreases to $\sim25\%$ if we consider individual resolved nuclei. The LIRG CONs, on the other hand, are often found in seemingly isolated galaxies with relatively settled morphologies, although some of them, for example IRAS~17578-0400 \citep{sti13} and NGC~4418 \citep{boe20}, do have nearby companions. In the latter case, the central gas concentration may be explained by a minor interaction with the companion \citep{boe20}, but the presence of CONs in isolated and undisturbed galaxies raises questions on the origin of the massive amounts of gas and dust in their nuclei, which in turn should affect the growth of the SMBH. It is possible, and maybe even likely, that there are multiple formation processes where the ULIRG CONs are formed in major mergers, while others form predominantly through minor interactions or secular processes internal to the galaxies. This would be similar to, for example, how pseudobulges are thought to form through secular evolution while classical bulges are formed through major mergers \citep[e.g.,][]{kor04}. Detailed studies of CONs, their host galaxies, and the environments of the hosts will be required to determine whether CONs are always formed as a result of interactions or if secular evolution is capable of depositing enough gas in the nuclei for CONs to form.

\subsubsection{Effects of the selection bias}\label{sec:effects_bias}
The difference in the distributions of the $f_{25}/f_{60}$ ratios in the CON and non-CON parts of the sample calls for a discussion of how the selection bias presented in Sect. \ref{sec:bias} affects our conclusions. The main effect of the bias is to prevent the real fraction of warmer sources that host CONs to be determined. Depending on the value of this fraction there are two main scenarios: either NGC~4418 is an outlier, and in general warm CONs are rare; or there is a second population of warm CONs. In order to discern between the two scenarios, a sample selected on the total infrared luminosity could be observed. This would include an additional six ULIRGs, $24$ LIRGs, and four sub-LIRGs compared to the current CON-quest sample. In the first scenario mentioned above, no additional CONs would be found in this extended sample and the fraction of (U)LIRGs that host CONs would have to be revised down by $\sim 50\%$ as the ULIRG and LIRG samples approximately double. In the second scenario, the detection rate of new CONs would depend on the actual fraction of warm galaxies that host CONs.
        
\subsection{Do all compact obscured nuclei have in- and outflows?}
The four previously known CONs in this survey were all included in a study of OH outflows in obscured galaxies \citep{fal19}. Contrary to most other (U)LIRGs in that study, the CON hosts all had positive median velocities, indicating inflowing motion, in the OH $119$~$\mu$m absorption lines. One other CON, IRAS~F22491-1808, was included in the same study, but was just below the limit to be considered a CON with the definition based on the $L_{\mathrm{HCN-vib}}/L_{\mathrm{IR}}$ ratio. Its OH $119$~$\mu$m lines are redshifted as in other CONs. One of the other newly identified CONs in this survey, ESO~320-G030, has observations of the OH $119$~$\mu$m doublet reported by \citet{gon17} under the name IRAS~11506-3851. Similarly to those in the already known CONs, the $119$~$\mu$m absorption lines of OH in ESO~320-G030 show evidence of inflowing gas in the form of a significant redshift relative to the systemic velocity of the galaxy. The other newly identified CON, IRAS~17578-0400, had an observation of the other ground state doublet at $79$~$\mu$m, with the absorption lines peaking at a slightly redshifted velocity (see Appendix \ref{app:OH}). In a multitransition OH analysis of molecular outflows by \citet{gon17} it is seen (their Fig. 10) that all sources in which the $79$~$\mu$m doublet peaks at velocities ${\gtrsim}0$~km\,s$^{-1}$ also have OH $119$~$\mu$m lines which peak at positive velocities.

At (sub)millimeter wavelengths, all seven CONs in our survey show possible signatures of molecular outflows. At least three of these, those in Arp~220 W \citep{bar18}, ESO 320-G030 \citep{per16}, and Zw~049.057 \citep{fal18}, appear collimated. This may be true also for the outflow in IRAS~17208-0014 \citep{gar15}, although its morphology still requires further study. IRAS~F22491-1808 has outflow signatures on kpc scales in the $J=2\text{--}1$ transition of CO reported by \citet{per18}. In NGC~4418, a compact molecular outflow detected in low-$J$ transitions of CO has been reported by \citet{flu19} and \citet{lut20} but its geometry is not known. The final CON, IRAS~17578-0400, does not have any published outflow signatures, but our HCN observations reveal a low-velocity elongation along its kinematic minor axis (Fig. \ref{fig:outflow_I17578}). This may be a signature of a minor axis outflow which is directed almost perpendicular to our line of sight toward the galaxy. Possible signatures of noncircular motions are also seen in the ground state HCN and HCO$^{+}$ lines in the form of asymmetric line profiles in some CONs. In most of them, the absorptions are stronger on the blueshifted side: Zw~049.057 \citep{aal15b}, IRAS~17208-0014 (Fig. \ref{fig:detections}), and Arp~220 \citep{sak09,aal15b,mar16} or close to the line center: IRAS~17578-0400 (Fig. \ref{fig:detections}), and NGC~4418 \citep{sak13}. In ESO~320-G030 (Fig. \ref{fig:detections}), however, it is clearly stronger on the redshifted side, possibly indicating that the absorbing gas is moving toward the nucleus, consistent with what is seen in CO $J=2\text{--}1$ observations by \citet{gon21}. We note, however, that these asymmetries may also be due to lopsided distributions of material in the nucleus rather than noncircular motions. Furthermore, noncircular motions do not necessarily indicate in- or outflows.

An interesting object for comparison is the CON IC\,860, which is not in our sample due to its declination ($+24\degr$) but has been studied at high spatial resolution by \citet{aal19}. In this galaxy, the $J=1\text{--}0$ line of CO observed by \citet{lut20} reveals no clear outflow signatures. However, \citet{aal19} observe blueshifted absorption in the CS $J=7\text{--}6$ transition toward the central region. They present two scenarios for the morphology of the galaxy: one with a near face-on disk and an outflow oriented approximately along our line of sight, and another where the disk is instead highly inclined and the outflow is closer to the plane of the sky. Although they could not distinguish between the two scenarios, \citet{aal19} note that the first scenario requires additional foreground components that the second one can do without. In the second scenario, the combination of a highly inclined disk with a minor axis outflow results in a structure that appears more face-on than it really is. This is similar to the situation suggested by \citet{sak17} to explain the CO velocity field in Arp~220. The effect may also be present in other CONs, for example ESO~320-G030 and IRAS~17578-0400 where the nuclear HCN emission is more extended along the kinematic minor axis than along the kinematic major axis (see Figs. \ref{fig:outflow_ESO320} and \ref{fig:outflow_I17578}).

\subsection{Alternative tracers of compact obscured nuclei}
\begin{figure}
  \centering
    \includegraphics[width=\hsize]{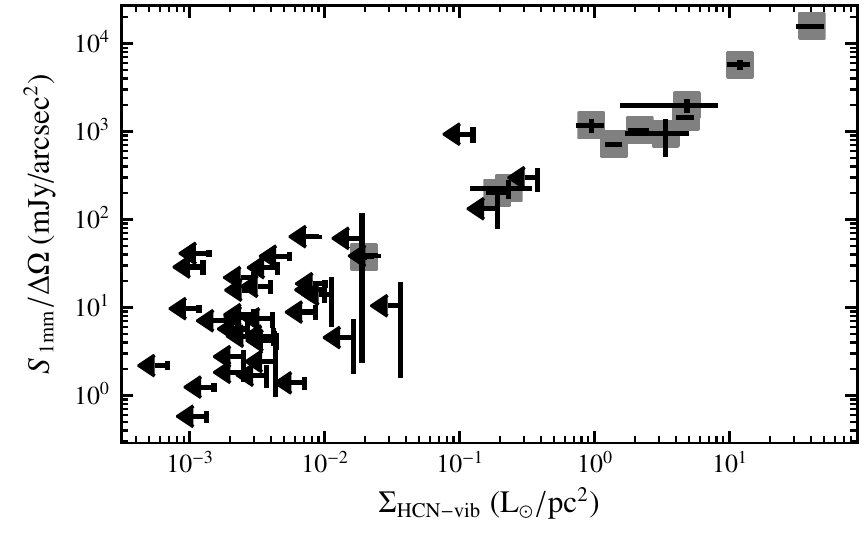}
  \caption{Continuum surface brightness at $1.1$~mm as a function of the HCN-vib surface brightness of the CON-quest sample.}
         \label{fig:HCNvib_vs_1mm}
\end{figure}

Determining the HCN-vib surface brightness requires observations with both high sensitivity and spatial resolution. It is not always possible to achieve both of these, so alternative ways to find CONs are sometimes desirable. With high enough spatial resolution, one possible way to find sources with very high column density is to examine the continuum brightness at (sub)millimeter wavelengths, where the optical depth of the dust approaches unity for column densities of $N_{\mathrm{H2}}\sim 10^{25}$~cm$^{-2}$. This method has been used in, for example, Arp~220 \citep{sak08}, NGC~4418 \citep{sak13}, and IC~860 \citep{aal19}. In Fig. \ref{fig:HCNvib_vs_1mm} we plot the continuum surface brightness at $1.1$~mm as a function of the HCN-vib surface brightness for all sources with HCN-vib detections. Among the HCN-vib detected sources there is a strong correlation, with a Pearson correlation coefficient of $>0.99$. We note that the surface brightnesses on both axes are calculated using the same area, that of the $1.1$~mm continuum, effectively meaning that brighter $1.1$~mm sources are also brighter HCN-vib sources. In most cases it is possible to use the continuum surface brightness to find CONs. However, one caveat is that the millimeter continuum may also be affected by synchrotron or free-free emission \citep[e.g.,][]{aal19}, a problem that gets worse at longer wavelengths. This method may thus render some false positives when searching for CONs. For example, NGC~1068 has a continuum surface brightness at $1.1$~mm of ${\sim} 1000$~mJy\,arcsec$^{-2}$, similar to many CONs, but a HCN-vib surface brightness of less than $0.13$~$L_{\sun}$\,pc$^{-2}$.

\begin{figure}
  \centering
    \includegraphics[width=\hsize]{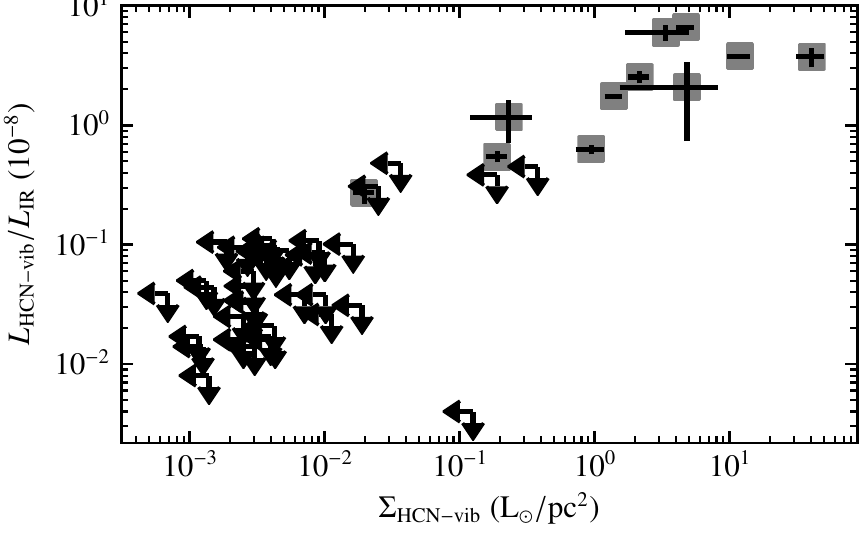}
  \caption{Luminosity of HCN-vib relative to the total (between $8$ and $1000$~$\mu$m) infrared luminosity as a function of the HCN-vib surface brightness for the CON-quest sample.}
         \label{fig:HCNvib_ratio_vs_density}
\end{figure}

In some cases, for example at higher redshift ($z\gtrsim0.15$), the spatial resolution will not be high enough to derive surface brightnesses. Although not completely equivalent, we saw in Sect. \ref{sec:detection_rate} that the criterion $L_{\mathrm{HCN-vib}}/L_{\mathrm{IR}}>10^{-8}$ selects mostly the same sources as the currently used definition and that it therefore may be useful in such situations. In Fig. \ref{fig:HCNvib_ratio_vs_density} we plot the $L_{\mathrm{HCN-vib}}/L_{\mathrm{IR}}$ ratio as a function of the HCN-vib surface brightness. The correlation between these quantities is weaker with a Pearson correlation coefficient of ${\sim}0.3$. This can be considered a moderate correlation and the criterion based on $L_{\mathrm{HCN-vib}}/L_{\mathrm{IR}}$ ratio may be used when the surface brightness is hard to derive.

\section{Conclusions}\label{sec:conclusions}
We present the first results of a systematic survey of infrared luminous galaxies, called CON-quest, with the aim to search for compact obscured nuclei, traced by strong HCN-vib emission. Our sample consists of literature, archival, and new ALMA data toward $46$ far-infrared selected ULIRGs ($10^{12}$~L$_{\sun} \leq L_{\mathrm{FIR}}$), LIRGs ($10^{11}$~L$_{\sun} \leq L_{\mathrm{FIR}} < 10^{12}$~L$_{\sun}$), and sub-LIRGs ($10^{10}$~L$_{\sun} \leq L_{\mathrm{FIR}}<10^{11}$~L$_{\sun}$).

Using our definition of a CON as a galaxy where $\Sigma_{\mathrm{HCN-vib}}> 1$~L$_{\sun}$~pc$^{-2}$ over nuclear regions with radii of ${\sim}5-100$~pc, we find that ${\sim}40\%$ of the ULIRGs, ${\sim}20\%$ of the LIRGs, and $<9\%$ of the sub-LIRGs, host CONs. It has been suggested that the CON phenomenon is related to inclination, but there is no evidence that the CON hosts have a different distribution of inclinations than the rest of the sample. However, the disk of the host galaxy may not be aligned with that of the nucleus, and high-resolution observations should be conducted to determine the inclinations of the nuclear disks in CONs. One property that does differ between CONs and other galaxies is the IRAS $f_{25}/f_{60}$ ratio, with CONs generally having lower ratios indicating colder dust spectral energy distributions (SEDs). This is consistent with the notion that the large dust columns in CONs work to gradually shift the radiation to longer wavelengths, making the mid- and far-infrared ``photospheres'' significantly cooler than the interior regions. There is however one outlier, NGC\,4418, with a significantly higher $f_{25}/f_{60}$ ratio which leaves open the possibility of a broad or bimodal $f_{25}/f_{60}$ distribution among CONs. The obscuration of the warm interiors by cooler foreground gas may also explain why the mid-infrared diagnostics EQW$_{6.2\mathrm{\mu m}}$ and $s_{9.7\mathrm{\mu m}}$ indicate that some of the deeply obscured CONs are instead relatively unobscured starbursts. So far, all CONs have possible signatures of inflowing molecular gas in the far-infrared. At (sub)millimeter wavelengths, however, all CONs show possible signatures of compact and collimated molecular outflows. Further high-resolution observations should be conducted to investigate if and how these properties are connected to each other and to the CON phenomenon. Detailed studies of individual sources, including for example multifrequency continuum observations \citep[e.g.,][]{aal19} or excitation analysis of HCN-vib at frequencies less affected by dust obscuration \citep[e.g.,][]{sal08}, would also make it possible to better constrain physical properties such as column densities, sizes, and temperatures of the nuclei.

The fact that our sample is based on far-infrared luminosities means that it is biased toward sources with cooler SEDs. The bias mainly affects the LIRG and ULIRG parts of the sample and can be remedied by obtaining ALMA measurements of the HCN-emission in a complementary sample of sources selected based on their total infrared luminosity. This would approximately double the number of LIRGs and ULIRGs in our sample and increase the sub-LIRG sample by $\sim 20\%$. If no more CONs are found in this extended sample, it would mean that the fraction of (U)LIRGs that host CONs has to be revised down by $\sim 50\%$. Regardless of whether more CONs with a warm SED are found or not, detailed observations of individual sources are required to determine if, and possibly how, cold and hot CONs differ. Finally, to make an assessment of the importance of CONs in the context of galaxy evolution, it is necessary to conduct studies at higher redshifts where (U)LIRGs are more common.

\begin{acknowledgements}
  This paper makes use of the following ALMA data: ADS/JAO.ALMA\#2018.1.01344.S., ADS/JAO.ALMA\#2015.1.00404.S, ADS/JAO.ALMA\#2017.1.00082.S, ADS/JAO.ALMA\#2017.1.00057.S, ADS/JAO.ALMA\#2013.1.00032.S, ADS/JAO.ALMA\#2017.1.00598.S, ADS/JAO.ALMA\#2015.1.00412.S.  ALMA is a partnership of ESO (representing its member states), NSF (USA) and NINS (Japan), together with NRC (Canada), MOST and ASIAA (Taiwan), and KASI (Republic of Korea), in cooperation with the Republic of Chile. The Joint ALMA Observatory is operated by ESO, AUI/NRAO and NAOJ.
  We thank the anonymous referee for useful comments and suggestions.
  We acknowledge support from the Nordic ALMA Regional Centre (ARC) node based at Onsala Space Observatory. The Nordic ARC node is funded through Swedish Research Council grant No 2017-00648.
  S.A. gratefully acknowledges support from an ERC AdvancedGrant 789410 and from the Swedish Research Council. K.S.
  S.G-B. acknowledges support from the research projects PGC2018-094671-B-I00 (MCIU/AEI/FEDER, UE) and PID2019-106027GA-C44 from the Spanish Ministerio de Ciencia e Innovaci\'on
  T.D-S. acknowledges support from the CASSACA and CONICYT fund CAS-CONICYT Call 2018.
  G.A.F acknowledges financial support from the State Agency for Research of the Spanish MCIU through the AYA2017-84390-C2-1-R grant (co-funded by FEDER) and through the “Center of Excellence Severo Ochoa” award for the Instituto de Astrof\'isica de Andalucia (SEV-2017-0709).
  T.R.G. acknowledges the Cosmic Dawn Center of Excellence funded by the Danish National Research Foundation under grant No. 140.
  E.G-A. is a Research Associate at the Harvard-Smithsonian Center for Astrophysics, and thanks the Spanish Ministerio de Econom\'{\i}a y Competitividad for support under projects ESP2017-86582-C4-1-R and PID2019-105552RB-C41.
  This research has made use of NASA's Astrophysics Data System.
  This research has made use of the NASA/IPAC Extragalactic Database (NED) which is operated by the Jet Propulsion Laboratory, California Institute of Technology, under contract with the National Aeronautics and Space Administration.

\end{acknowledgements}

\bibliographystyle{bibtex/aa} 
\bibliography{ref} 

\begin{appendix}
  \section{Herschel OH detection in IRAS 17578-0400}\label{app:OH}
\begin{figure}[ht]
   \centering
   \includegraphics{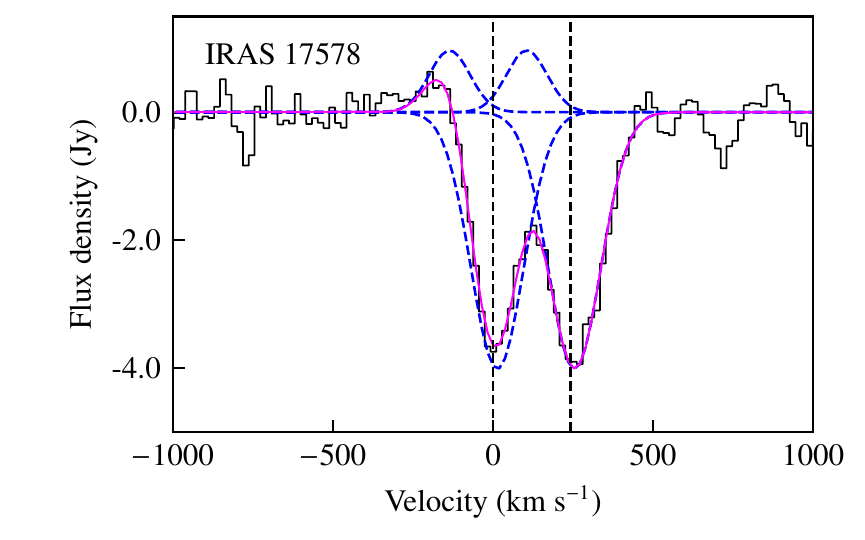}
   \caption{Spectral fits to the OH $79$~$\mu$m absorption lines in IRAS~17578-0400. The solid black histogram represents the data, the solid magenta line is the best multicomponent fit to the data, and the dashed blue lines are the individual components. The velocity scale is set relative to the frequency of the blue component of the doublet. Dashed vertical lines indicate the expected positions of the two absorption components given the adopted redshift of $0.0134$.}
              \label{fig:new_oh}%
  \end{figure}

  One of the newly identified CONs, IRAS 17578-0400, also had observations of the OH ground state doublet at $79$~$\mu$m taken with the Photodetector Array Camera and Spectrometer \citep[PACS;][]{pog10} on the \emph{Herschel} Space Observatory \citep{pil10}. The observations (obsid: 1342239716, PI: L.~Armus) were conducted on 2012 February 25 with the high spectral sampling, range spectroscopy mode for a duration of $1176$~s and the data were processed with version 14.2 of the standard pipeline. The nuclear far-infrared emission of IRAS~17578-0400 is spatially unresolved in the central $9.4$\arcsec\ (${\sim}3$~kpc) spaxel of the PACS $5$~x~$5$ spaxel array. As the central PACS spaxel is smaller than the point spread function of the spectrometer, the spectrum was extracted using the point source correction task in the \emph{Herschel} interactive processing environment \citep[HIPE;][]{ott10} version 14.0.1. Before analyzing the absorption lines, a polynomial of order two was fitted to the continuum and then subtracted from the spectrum. To ease comparison with the OH $119$~$\mu$m doublets used in previous studies, the profiles of the OH $79$~$\mu$m doublets were modeled using the same procedure as in \citet{vei13}. Each line was fitted with two Gaussian components characterized by their amplitude, velocity centroid, and width. The separation between the two lines of the doublet was fixed at $0.064$~$\mu$m in the rest frame and the amplitude and width of the two lines were the same for each component. The median velocity of the absorptions was then found to be $33$~km\,s$^{-1}$. The fitting procedure was carried out using the spectroscopic analysis toolkit \emph{PySpecKit} \citep{gin11} and the continuum subtracted spectrum with the fits overplotted are presented in Fig. \ref{fig:new_oh}. 

\end{appendix}

\end{document}